\documentclass[twocolumn,epjc3]{svjour3}          % twocolumn

\RequirePackage[T1]{fontenc}

\smartqed  % flush right qed marks, e.g. at end of proof
\usepackage{packages}
\RequirePackage{graphicx}
\RequirePackage{mathptmx}      % use Times fonts if available on your TeX system
\RequirePackage{flushend}
\RequirePackage[numbers,sort&compress]{natbib}
\RequirePackage[colorlinks,citecolor=blue,urlcolor=blue,linkcolor=blue]{hyperref}

\journalname{Eur. Phys. J. C}

\begin{document}

\title{Hadronic heavy neutral lepton decays to the limit}

% \subtitle{ } 

\author{
        Jonathan L. Schubert\thanksref{e1,mpp,tum}
        \and
        Babette D{\"obrich}\thanksref{e3,mpp} 
}

\thankstext{e1}{e-mail: jonathan.schubert@cern.ch}
\thankstext{e3}{e-mail: babette@mpp.mpg.de}

\institute{
            Max-Planck-Institut für Physik (Werner-Heisenberg-Institut), Boltzmannstr. 8, 85748 Garching bei München, Germany\label{mpp}
          \and
            Technical University of Munich, TUM School of Natural Sciences, Physics Department, Chair for Data Science in Physics, 85748 Garching, Germany\label{tum}
}

\date{September 2026 }
% The correct dates will be entered by the editor

\maketitle

\begin{abstract}
Heavy Neutral Leptons are a class of hypothetical particle, motivated by simultaneously solving multiple of the standing issues of the Standard Model.
In this work we investigate the decay structure of this type of particle in the hadronic picture for multi-body final states.
We propose to link the resulting hadronic description to a partonic description in a seamless transition based on the hadron system's invariant mass.
Beyond the impact on the total HNL lifetime, we will show that the multi-hadron channels emerging in this description can constitute interesting experimental signatures. 
\end{abstract}
\section{Introduction}

The search for long-lived particles (LLPs) with masses in the MeV to GeV range has become one of the most active frontiers in the accelerator-based program aiming to find physics beyond the Standard Model. 
While much of the activity had traditionally focused on the electroweak and TeV scales, increasing attention has been given to the MeV-GeV scale: This is due to to a broad class of motivated models that evade `conventional' collider searches owing to their tiny interactions and macroscopic lifetimes, see e.g. Ref.s~\cite{Batell:2022dpx,PBC:2025sny,Alimena:2025kjv} for recent reviews on the topic.
One of the arguably most motivated such models is that of a heavy neutral lepton (HNL) owing to their role in explaining neutrino masses, the baryon asymmetry of the Universe, and maybe also featuring a viable dark matter candidate.
From an experimentalist point of view, HNLs have a special role thanks to the diversity of their decay signatures. 
By construction, through their mixing with the Standard Model neutrinos, HNLs inherit weak interactions and can decay through charged- (CC) and neu\-tral-current (NC) processes into a wide range of final states. 
These include purely leptonic channels and  semileptonic decays involving, depending on the scale in the decay, meta\-sta\-ble mesons or quarks.
However, the branching ratio for each of these channels as a function of HNL mass is non-trivial to predict, especially for HNLs at scales of a few GeV where the hadronic description transitions to the partonic picture.

In this mass regime, HNLs are commonly approximated in terms of a single HNL $N$ with phenomenological Lagrangian~\cite{Gorbunov:2007ak,Atre:2009rg,Helo:2010cw,Bondarenko:2018ptm,Coloma:2020lgy,Feng:2024zfe}
\begin{equation}\label{eq:Lagrangian_HNL_int}
    \begin{split}
    \mathcal L _\mathrm{pheno} = &\frac g {2\sqrt 2} W_\mu^+\sum_{\alpha} \theta_{\alpha}^* \bar{N^c}\gamma^\mu (1-\gamma_5) \ell^-_\alpha \\
    &+ \frac g {4\cos\theta_W}Z_\mu \sum_{\alpha} \theta_{\alpha}^* \bar{N^c}\gamma^\mu (1-\gamma_5) \nu_\alpha + h.c.
    \end{split}
\end{equation}
where the interactions with lepton families $\alpha$ are suppressed by the mixing angles $\theta_{\alpha}$.
% \begin{equation}
%     \theta_{\alpha} \equiv \frac{F_{\alpha I }v}{\sqrt 2 m_N}.
% \end{equation}
It is further convenient to define the coupling suppression $U_\alpha^2 = |\theta_\alpha|^2$, which in the following we will often refer to simply as the coupling.\footnote{All quantities in this work are calculated for a Dirac HNL. 
The corresponding widths of Majorana HNLs, for which the hermitian conjugate processes are allowed, can be derived by applying a factor of 2~\cite{Bondarenko:2018ptm}.
Branching ratios are unaffected by the nature of the HNL, notwithstanding charges of the final state particles.}

The decays of such HNLs into hadrons are a broadly covered topic in literature~\cite{Gorbunov:2007ak,Atre:2009rg,Helo:2010cw,Bondarenko:2018ptm,Coloma:2020lgy,Feng:2024zfe}.
However, they are commonly considered through the lens of effective two-body decays.
In a frequently applied prescription, heavier HNLs decay into increasingly wide resonances, while multi-body final states are estimated from the difference between hadronic and partonic width evaluated for a given mass $m_N$~\cite{Coloma:2020lgy,Feng:2024zfe}, or relying on a fully partonic description that hadronises above a given threshold $m_N$, usually chosen to be about 1\,GeV.~\cite{Bondarenko:2018ptm,SensCalc}
A recent work on QCD corrections to CC-mediated HNL decay widths provides an alternative approach, treating these HNL decays in a fully inclusive way.~\cite{Kretz:2025pfk}

In this work, we aim to propose a different prescription, relying on multi-body hadronic decays.
To the our know\-ledge, Bondarenko et al.~\cite{Bondarenko:2018ptm} are the only work on the subject presenting a full three-body treatment for hadronic HNL decays, albeit limited to $N\to\ell\pi\pi^0$ and $N\to\nu\pi\pi$ decays.\footnote{The authors of the paper discuss the full treatment in the context of estimating sensitivity for counting experiments, leading them to the conclusion that the phenomenology is sufficiently captured by the effective decay description with an intermediate on-shell $\rho$. 
Also, decays into an on-shell $a_1$ resonance are discussed qualitatively, but to our knowledge, no quantitative results are presented.}
The fundamental idea of the change in description boils down to identifying a scale, where the the hadronic transitions into the partonic picture.
A similar approach has been taken for other Long Lived Particle Scenarios (e.g. gluon-quark-coupled axion-like-particles\cite{Balkin:2025enj} and dark scalars~\cite{Blackstone:2024ouf,Gieseke:2025gfq}).
In these LLP scenarios, decays are largely either leptonic or hadronic.
Therefore, such a scale is easily associated with the LLP mass.
In the HNL case, the semi-leptonic nature of the decays means that the hadronic scale of a decay is only indirectly related to the the HNL mass. 
As we will see below, it is nonetheless possible to identify a suitable projection (akin to the well known spectral distribution), where the transition scale can be identified.

We evaluate the partial widths presented in this work using the \textsc{Alpinist} framework~\cite{Jerhot:2022chi}, building on the implementation of relevant production and decay mechanisms of HNLs in the GeV regime~\cite{Schubert:2024hpm, Alpinist_Zenodo}.
\textsc{Alpinist} is a public toy-MC framework that enables us to employ a common set of theory assumptions for LLP production for a wide set of experiments, originally conceived for axion-like-particles at proton-beam-dumps. 
Notably, for the purpose of this study, we expand the implementation by revising the HNL implementation of resonant hadronic contributions and allowing the hadronisation of partonic states.
The largely theory-independent MC sensitivity estimation within \textsc{Alpinist} then allows us to showcase the sensitivity impact of different assumptions on semi-leptonic HNL decays.

The paper is structured as follows: We first review the phenomenology of hadronic HNL decay channels in two- and three-body decays. 
This is complemented by a general description of three-body decays involving pseudo-scalar final states and resonant four-body decays involving three pions in Sect.~\ref{sec:hadronic_decays}.
Extended calculations and literature are provided in~\ref{sec:decay_pars}. 
We will then introduce the partonic picture of semileptonic HNL decays in Sect.~\ref{sec:quark_width}.
Following these reviews, Sect.~\ref{sec:transition} will illustrate ways of quantifying the duality of these two descriptions, which will allow us to find a natural transition point from hadronic to partonic scale.
The resulting prescription yields differences with respect to the literature estimation of branching ratios and lifetimes of HNL masses with masses in the 1 to 10\,GeV range at the $10\,$\% level.
Finally, we will discuss the experimental implications of this prescription before concluding.

\section{Hadronic decay channels of the heavy neutral lepton}
\label{sec:hadronic_decays}

\begin{figure}
    \centering
    % \resizebox{\linewidth}{!}{
    % \input{Plots/Feynman/HNL_CC_decay}
    % \input{Plots/Feynman/HNL_NC_decay}
    % }
    \includegraphics[width=0.495\linewidth]{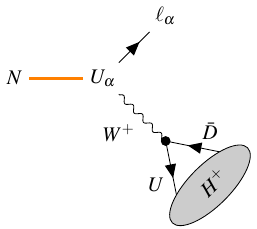}
    \includegraphics[width=0.495\linewidth]{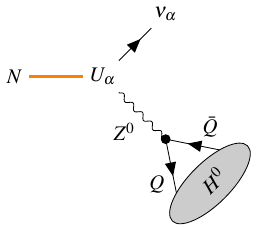}
    \caption{Schematic semi-leptonic decay of a heavy neutral lepton $N$ in a charged- (left) and neutral- (right) current-mediated transition into hadronic systems $H$.}
    \label{fig:schematic_HNL_decay}
\end{figure}

Semileptonic decays of the HNL occur in CC or NC transitions as schematically presented in Fig.~\ref{fig:schematic_HNL_decay}.
In the CC case, the $W$ mediator  couples to an up-anti-down-type quark-anti\-quark pair $U\bar{D}$ which showers and hadronises into a charged possibly multi-body hadronic system $H^+$.
In the NC case, the equivalent process occurs via a $Z$ boson into an intermediate quark-antiquark pair $Q\bar{Q}$ pair hadronising into a neutral system $H^0$.
In the low energy regime it is fruitful to treat these structures as HNL decays directly into hadronic final states.

The amplitudes of these processes are described as 
\begin{equation}
    \frac{G_FU_\ell }{\sqrt{2}} \underbrace{\bar u_N \gamma^\mu(1-\gamma_5) u_{\ell/\nu}}_{L^\mu}\, n_J\underbrace{\langle 0|j_\mu|H\rangle}_{J_\mu},
\end{equation}
where $u_N$ and $u_{\ell/\nu}$ are the HNL and lepton/neutrino spi\-nors, $G_F$ is the Fermi coupling constant, and $j^\mu$ the hadronic transition current relating the hadron system $H$ to the vacuum.
In CC transitions, the hadronic current normalisation $n_J$ is the Ca\-bib\-bo–Ko\-ba\-ya\-shi–Ma\-ska\-wa matrix element $V_\mathrm{UD}$ applicable to the underlying quark transition between up-like ($U$) and down-like ($D$) quarks.
In the NC case,  $n_J=1$.
These structures can be grouped as the leptonic ($L^\mu$), and hadronic current ($J^\mu$).

As the description of multi-hadron decays largely relates to current exchange structures known from $\tau$ decays, the expressions for $J_\mu$ are readily derived.
In this section we will outline the relevant formulas describing HNL decays in the hadronic picture from two- to four-body decays.

\subsection{Single hadron}\label{sec:two_body_decays}

The decay width to neutral and charged pseudo-scalar mesons $h_P$ is given by~\cite{Gorbunov:2007ak,Bondarenko:2018ptm, Coloma:2020lgy}
\begin{equation}\label{eq:twobody_pseudoscalar_decays}
    \Gamma_{N\to\nu_\alpha h_P^0}=\frac{G_F^2f_h^2U_\alpha^2m_N^3}{32\pi}(1-x_h^2)^2\text{, and}
\end{equation}
\begin{align}
\begin{split}
    \Gamma_{N\to\ell_\alpha^- h_P^+}=&\frac{G_F^2f_h^2|V_{UD}|^2 U_\alpha^2m_N^3}{16\pi}\sqrt{\lambda(1,x_\ell^2,x_h^2)}\\
    &\quad\times\Big[(1-x_\ell^2)^2-x_h^2(1+x_\ell^2)
    \Big]
\end{split} \end{align}
respectively. 
Here, $x_h=m_h/m_N$, $x_\ell=m_\ell/m_N$, $f_h$ is the pseu\-do-scalar's decay constant as given in Tables~\ref{tab:charged_decay_constants} and~\ref{tab:neutral_decay_constants}, and  $\lambda$ is the K{\"a}ll{\'e}n function
\begin{equation}
    \lambda(a,b,c) = a^2+b^2+c^2-2ab-2bc -2ca.
\end{equation}
Using the same notation, the decay width to neutral and vector mesons $h_V$ is given by~\cite{Atre:2009rg, Bondarenko:2018ptm, Coloma:2020lgy}
\begin{align}\label{eq:}
    \Gamma_{N\to\nu_\alpha h_V^0}=&\frac{G_F^2\kappa^2_{h}g_h^2U_\alpha^2m_N^3}{32\pi m_h^2}(1+2x_h^2)(1-x_h^2)^2,\,\text{ and}\\
    \begin{split}
    \Gamma_{N\to\ell^-_\alpha h_V^+}=&\frac{G_F^2g_h^2 |V_{UD}|^2U_\alpha^2m_N^3}{16\pi m_h^2}\Big[(1-2x_\ell^2)^2\\
    &\qquad+x_h^2(1+x_\ell^2-2x_h^2)\Big]\sqrt{\lambda(1,x_\ell^2,x_h^2))},
    \end{split}
\end{align}
where $\kappa_h$ is the proportionality constant relating the vector meson current to the the neutral current, and $g_h$ is the vector meson decay constant (for details and discussion see~\ref{sec:hadronic_two_body_pars}) with values given in Tables~\ref{tab:neutral_vector_decay_constants} and~\ref{tab:charged_vector_decay_constants}.

Discrepancies in terms of prefactors of 2 for the neutral current-mediated two-body decays in Ref.s~\cite{Atre:2009rg,Helo:2010cw} have been discussed in Ref.s~\cite{Bondarenko:2018ptm,Coloma:2020lgy}.
We find agreement with Ref.s~\cite{Bondarenko:2018ptm,Coloma:2020lgy} in both CC and NC two-body decays.
The numerical values of $\kappa_h$ for $\omega$, and $\phi$ are the main difference between Ref.s~\cite{Bondarenko:2018ptm} and~\cite{Coloma:2020lgy}, where we agree with~\cite{Coloma:2020lgy}.
Equivalently, this discrepancy extends to $\kappa_{J/\psi}$.
A point of difference with respect to Ref.s~\cite{Bondarenko:2018ptm,Coloma:2020lgy} is our treatment of the effective $\eta$ and $\eta^\prime$ decay constants, where we largely agree with values~\cite{Feldmann:1999uf} cited by Ref.~\cite{Atre:2009rg}.
For further details find \ref{sec:hadronic_two_body_pars}.

\subsection{Two hadrons}

Three-body decays into pseudo-scalar mesons are of the form $N\to\ell h^+_1 \overline{h^0_2}$, or $N\to\nu h \overline{h}$.
The most general expression of the charged hadronic current-mediating a lepton decay into two pseudo-scalar mesons $h_1$ and $h_2$ is given by~\cite{Finkemeier:1996dh,Bernard:2011ae}
\begin{equation}\label{eq:2hadron_current}
\begin{split}
   J^\mu=&F^V(Q^2)\left((p_{h2}-p_{h1})^\mu -\frac{m_{h2}^2-m_{h1}^2}{Q^2}(p_{h2}+p_{h1})^\mu\right)\\
   &+F^S(Q^2)\left(\frac{m_{h2}^2-m_{h1}^2}{Q^2}(p_{h2}+p_{h1})^\mu\right),
\end{split}
\end{equation}
where $Q= p_{h2}+p_{h1}$, and  $F^V(Q^2)$ and $F^S(Q^2)$ are the vector and scalar form factors, respectively.%\footnote{In this convenient convention for the form factors, the normalisations $F^V(0)$ and $F^S(0)$ are the same.} 
The part of this current proportional to $(m_{h2}^2-m_{h1}^2)/Q^2$ manifestly leads to expressions proportional to the squared mass difference of the mesons and can, thus, be neglected when deriving expressions for meson pairs.
However, including the full description, including mass splitting, is crucial for describing, for example, the $N\to\ell (K^*\to K\pi)$ decay. 

Writing $F^{V/S}=F^{V/S}(Q^2)=F^{V/S}(\xi m_N^2)$ for better readability, we find 
\begin{equation}\label{eq:N_ellh1h2}
\begin{split}
    &\Gamma_{N \to \ell_\alpha^- h_1^+\bar{h}^0_2 }= \frac{G_F^2 m_N^5}{768 \pi^3}\left|V_{UD}\right|^2 U_\alpha^2 \\
    &\quad\times\int\limits_{4 x_h^2}^{\left(1-x_{\ell}\right)^2}\mathrm d\xi \sqrt{\lambda\left(1, \xi, x_{\ell}^2\right)}\tilde\beta_{h,\delta}(\xi) \\
    &\quad\times\Bigg[ \left|F^V\right|^2 \tilde\beta^2_{h,\delta}(\xi)\left(\left(1-x_\ell^2\right)^2+\xi\left(1+x_{\ell}^2\right)-2 \xi^2\right)\\
    &\quad\qquad + 12\left|F^S\right|^2\frac{\delta_h^2x_h^2}{\xi^2}\left( (1-x_\ell)^2- \xi(1+x_\ell^2) \right)\Bigg],\,
\end{split}
\end{equation}
where $\delta_h=(m_{h2}-m_{h1})/m_N$, $x_{h}=(m_{h1}+m_{h2})/2m_N$, $\xi=Q^2/m_N^2$, and we defined
\begin{equation}
    \tilde\beta_{h,\delta}(\xi)=\sqrt{\left(1-\frac{4x_h^2}{\xi}\right)\left(1-\frac{\delta_h^2}{\xi}\right)}.
\end{equation}
In the limit of $x_\ell\to0$, this takes the known form for $\tau\to\nu_\tau K\pi$ decays~\cite{Finkemeier:1996dh}.

In the limit of $\delta_h\to0\Leftrightarrow m_{h1}=m_{h2}$, (ie. semi-leptonic decays of an HNL into a pair of pseudo-scalars), we recover the known expression~\cite{Bondarenko:2018ptm}.
\begin{align}\label{eq:N_ellhh}
\begin{split}
    &\Gamma_{N \to \ell_\alpha^- h^+h^0 }= \frac{G_F^2 m_N^5}{768 \pi^3}\left|V_{UD}\right|^2\left|U_\alpha\right|^2\\
    &\qquad\times\int\limits_{4 x_h^2}^{\left(1-x_{\ell}\right)^2}\mathrm d\xi\left(\left(1-x_\ell^2\right)^2+\xi\left(1+x_{\ell}^2\right)-2 \xi^2\right) \\
    &\qquad\qquad\times \sqrt{\lambda\left(1, \xi, x_{\ell}^2\right)} \tilde\beta_h^3\left(\xi\right)\left|F^V_{h}\left(\xi m_N^2\right)\right|^2. 
\end{split}  
\end{align}
Here again, $x_h=m_h/m_N$, $x_\ell=m_\ell/m_N$, $\tilde{\beta}_h(\xi)=\sqrt{1-4x_h/\xi}$, and $F_h^V$ is the CC vector form factor.
Differences in the prefactors stem from Ref.~\cite{Bondarenko:2018ptm} explicitly incorporating the $F_\pi^V$ proportionality to $F_\pi^{\mathrm{EM}V}$ into the width equations.

The NC-mediated decay can be described analogously, while $h_1 = h_2$ due to flavour conservation under NC.
With $F^{\mathrm{NC}V}_{h}$ as the NC vector form factor we find
\begin{align}
\begin{split}
    &\Gamma_{N \to \nu_\alpha h \bar h}=\frac{G_F^2 m_N^5}{768 \pi^3}\left|U_\alpha\right|^2 \\
    &\qquad \times \int\limits_{4 x_h^2}^1\mathrm d\xi\left(1-\xi\right)^2\left(1+2\xi\right) \tilde\beta_h^3\left(\xi\right)\left|F^{\mathrm{NC},V}_{h}\left(\xi m_N^2\right)\right|^2
\end{split}    
\end{align}
in agreement with the two-pion decay derived in Ref.~\cite{Bondarenko:2018ptm}.

The neutral and charged vector form factors can commonly be related by isospin arguments.
For the pion, they are further related to the electromagnetic form factor $F_\pi^{\mathrm{EM}V}$ as $F_\pi^{\mathrm{EM}V}=F_{\pi}^{\mathrm{CC}V}/\sqrt{2}=F_{\pi}^{\mathrm{NC}V}/(1-2 \sin ^2 \theta_W)$~\cite{Bondarenko:2018ptm}, which we pa\-ra\-me\-trise in a vector-meson-dominance (VMD) model using BaBar data~\cite{BaBar:2012bdw}.
In the Kaon case, $F_{K}^{\mathrm{NC}V}$ is a combination of isoscalar and isovector contributions, while the CC form factor $F_{K}^{\mathrm{CC}V}$ is purely isovector in nature.
We paramet\-rise $F_{K}^{\mathrm{CC}V}$ using an approach from dispersion theory as defined in Ref.~\cite{Gonzalez-Solis:2019iod} fit to BaBar data~\cite{BaBar:2018qry}, and extract the isoscalar component from an isoscalar-isovector deconstruct\-ed fit of the electromagnetic kaon form factors~\cite{Stamen:2022uqh}.
Vector and scalar form factors for $N\to K\pi\ell$ can be taken from fits to $\tau\to K \pi\nu$ and $K_{e3}$ transitions~\cite{Jamin:2006tk}.
We will use a VMD parametrisation of $\tau \to\nu K_S\pi^-$ decay events recorded by the Belle collaboration~\cite{Belle:2007goc}.
For further details on form-factor parametrisations see~\ref{sec:kaon_ff}.

\subsection{Three hadrons}
\label{sec:a1_resonance}

\begin{figure}
    \centering
    % \resizebox{\linewidth}{!}{\input{Plots/Feynman/tau3pi}\input{Plots/Feynman/hnl3pi}}
    \includegraphics[width=0.495\linewidth]{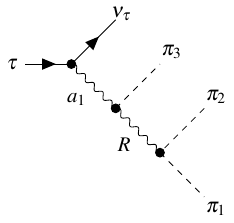}
    \includegraphics[width=0.495\linewidth]{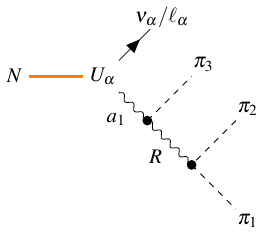}
    \caption{Decay to $3\pi$ final state mediated by the charged $a_1 \to R \pi$ resonance decay cascade of the $\tau$-lepton (left) and heavy neutral lepton (right). }
    \label{fig:a1_decay_structure}
\end{figure}

A class of processes commonly not fully appreciated in recent reviews of HNL GeV phenomenology (e.g.~\cite{Bondarenko:2018ptm, Coloma:2020lgy, Alimena:2025kjv}) is the decay into three light hadronic final states.
This is in spite of their significant branching fraction of the $\tau$ lepton, the decays of which are closely related to those of the HNL, as schematically shown in Fig.~\ref{fig:a1_decay_structure}.
In this work we limit ourselves to the case of three pions.

The decay can be modelled  as
\begin{equation}
    \mathrm d\Gamma(N 
    \to (\ell_\alpha/\nu_a) 3\pi)=\frac{4\pi^4U_\alpha^2\mathrm{n}_\mathrm{C}G_F^2}{m_N}\left|L^{\mu}J_\mu\right|^2\mathrm d \Phi_4\,,\label{eq:3pi_charged_amplitude}
\end{equation}
where $L^{\mu}$ is the leptonic current of the weak decay, $J_\mu$ is the hadronic weak current associated with $3\pi$ production, $\mathrm d \Phi_4$ is the four-body phase space, and $n_\mathrm{C}$ is the current normalisation.
In the CC transition, this is the the CKM element of the so that $n_\mathrm{CC}=|V_{ud}|^2$.
For a G-odd system, such as $3\pi$, the current structure is purely axial in the chiral limit~\cite{Kuhn:1990ad}, due to G-parity conservation.
For $\tau$ decays, this is in line with experimental data with upper limits on non-axial contributions at the 20\,\% level~\cite{PDG2026}.
Isospin symmetry leads to the relationship 
\begin{equation}
    \langle 3\pi|j^\mathrm{CC}_\mu|0\rangle \simeq \sqrt 2 \langle 2\pi\pi^0|j^\mathrm{NC}_\mu|0\rangle ,
\end{equation}
as argued in appx. B of Ref.\cite{Bondarenko:2018ptm}.
Here, $j^\mathrm{CC/NC}_\mu$ are the interaction currents of the charged and neutral weak interactions with an unflavoured quark system.
Consequently, we find $n_\mathrm{NC}=\tfrac{1}{2}$.

The possible axial vector resonances mediating $J_\mu$ are $a_1(1260)$ and its radial excitations~\cite{Kuhn:1990ad}.
In this work we will limit ourselves exclusively to $a_1(1260)$, as experimental insights on the impact of the radial excitations are an active area of research~\cite{COMPASS:2020yhb, Rabusov:2023tna}.
The hadronic current then takes the form~\cite{Kuhn:1990ad}
\begin{equation}
    J^\mu_{a_1}=-\frac{4}{3f_\pi}\mathrm{BW}_{a_1}(q^2_{3\pi})\sum_{R} A_R J^\mu_R\,,%\frac{m_{a_1}}{\Gamma^0_{a_1}}
\end{equation}
where $A_R$ are the complex amplitudes of the partial waves $J^\mu_R$ corresponding to a partial wave resonance $R$, and $BW_{a_1}$ is the dynamic $a_1$ Breit-Wigner (BW) propagator.
In this work, we limit ourselves to the $\rho$ meson as an intermediate resonance $R$ (with partial waves $S$ and $D$).
This is in line with dedicated $\tau$ decay Monte Carlo tools, where only the $\rho^{(\prime)}$ contribution is considered~\cite{Jadach:1990mz,Hagiwara:2012vz}.
We will assume BW parameters $m^0_{a1}=1.211$\,GeV and $\Gamma^0_{a1}=0.446$ as measured by ARGUS in the fully charged decay channel $\tau\to\nu3\pi$, compatible with current Particle Data Group (PDG) averages~\cite{ARGUS:1992olh, PDG2026}.
We will further assume an amplitude ratio $A^S_\rho/A^D_\rho=-0.11$ between the partial $S$ and $D$ waves as measured by ARGUS~\cite{ARGUS:1992olh}.
For a discussion of these assumptions and detailed modelling of the hadronic currents find~\ref{sec:a1_partial_waves}.

For the NC-mediated decays $N\to\nu\pi^+\pi^-\pi^0$ and $N\to\nu3\pi^0$, we employ the same partial waves and amplitudes as in the CC case. 
Consequently, the $N\to\nu3\pi^0$ is not included in the following calculations, as it cannot be mediated by the $\rho$ resonances due to G-parity conservation (forbidding $\rho^0\to\pi^0\pi^0$ decays).
Another consequence is that, as all final state particles in $N\to\nu\pi^+\pi^-\pi^0$ are distinct, Bose-symmetrisation of the amplitude is not required.
Nonetheless, the amplitude structures are related, as outlined in \ref{sec:a1_partial_waves}.

We evaluate the resulting (differential) partial widths by means of Monte Carlo integration.
Contributions to $N\to\nu\pi^+\pi^-\pi^0$ from $N\to\nu(\omega,\phi)$ and $N\to\nu3\pi^0$ from $N\to\nu\eta$ are considered in the two-body decay approximation of the decays.

\section{Partonic heavy neutral lepton width}\label{sec:quark_width}

At energies well above the hadron scale (but below the $W$ mass), the hadronic partial width of the HNL is described by its decays into quark pairs.
The partial decay widths to bare quarks of up-type $U$ or down-type $D$ can be expressed as~\cite{Gorbunov:2007ak,Atre:2009rg,Helo:2010cw,Bondarenko:2018ptm,Coloma:2020lgy}
\begin{equation}\label{eq:CC_3_body}
    \Gamma_{N\to \ell U\bar D}= \frac{m_N^5N_C|V_{UD}|^2G_F^2U_\alpha^2}{192\pi^3}I\left(x_\ell,x_D,x_U\right).
\end{equation}
Here, $N_C$ is the number of colours, $V_{UD}$ is the CKM matrix element connecting the $U$ and $D$ quarks, $x_\ell$, $x_U$, and $x_D$ are the ratios $m_\ell/m_N$, $m_U/m_N$, and $m_D/m_N$.
Integrating out the $UD$ system, we find the known expression~\cite{Gorbunov:2007ak,Bondarenko:2018ptm,Coloma:2020lgy}
\begin{equation}
\begin{split}\label{eq:G_uld}
  I(x_U,x_D,x_l) = 12&\int\limits_{(x_D +
    x_l)^2}^{(1-x_U)^2}\frac{\mathrm d\chi}{\chi}\sqrt{\lambda(\chi, x_\ell^2, x_D^2) \lambda(1, \chi, x_U^2)}\\
   &\quad\times\left( \chi - x_\ell^2 - x_D^2 \right)\left( 1 + x_U^2 - \chi \right),
\end{split}
\end{equation}
 where and $\chi$ is the ratio of the $D\ell$ system's invariant mass to the HNL mass squared ($q_{d\ell}^2/m_N^2$).
Importantly, even though $I$ is manifestly invariant under permutations of $x_U$, $x_D$ and $x_\ell$, the integrand of $I(x_\ell,x_U,x_D)$ does not correspond to the differential width with respect to the hadronic system. 
Integrating out the $D\ell$ system instead, we find
\begin{equation}
\begin{split}\label{eq:G_lud}&I\left(x_\ell,x_U,x_D\right)=\int\limits_{(x_U+x_D)^2}^{(1-x_\ell)^2}\frac{\mathrm d \xi}{\xi^3}\sqrt{\lambda\left(1,\xi,x_l^2 \right)\lambda\left(\xi,x_U^2,x_D^2\right)}\\
    &\qquad\times\Bigg[3\left((1-x_\ell^2)^2-\xi^2)(\xi^2-(x_U^2-x_D^2)^2\right)\\
    &\qquad\qquad-\lambda\left(1,\xi,x_l^2\right)\lambda\left(\xi,x_U^2,x_D^2\right)\Bigg],
\end{split}
\end{equation}
where now $\xi$ is the ratio of the quark system's invariant mass to the HNL mass squared ($q_H^2/m_N^2$).

For the equivalent NC-mediated decay into a quark-anti quark pair $Q\bar Q$, we find
\begin{align}
\begin{split}\label{eq:G_nuqq}
    &\Gamma_{N\to \nu Q\bar Q}= \frac{m_N^5N_CG_F^2U_\alpha^2}{192\pi^3}\int\limits_{4x_Q^2}^{1}\frac{\mathrm d \xi}{\xi}\sqrt{1-\frac{4x^2}{\xi}} (1-\xi)^2
    % \\&\quad\times\Bigg[2C_1^Q x (1+2 x)+ x_Q^2 \left(8C_2^Q (1-4 x)+(1-x)\right)\Bigg],
    \\&\quad\times\Bigg[2C_1^Q\left( \xi (1+2 \xi) + 2x^2(1-\xi)\right) + 24 C_2^Q \xi x^2\Bigg],
\end{split}
\end{align} 
where $x=m_Q/m_N$ and the constants $C_1$ and $C_2$ depend on the quark's electric charge. 
With $s_W$ as the sine of the Weinberg angle ($s_W^2=0.22305$) we find
\begin{align}
&C_1^U = \frac{1-\tfrac{8}{3} s_W^2+\tfrac{32}{9}s_W^4}{4}, &C_2^U = \frac{s_W^2 }{3} \left(\tfrac{4}{3} s_W^2-1\right), \\
&C_1^D = \frac{1-\tfrac{4}{3} s_W^2+\tfrac{8}{9}s_W^4}{4},  &C_2^D = \frac{s_W^2}{6}  \left(\tfrac{2}{3} s_W^2-1\right)
\end{align} 
for up(down)-type quarks $U$ ($D$) in agreement with~\cite{Bondarenko:2018ptm, Coloma:2020lgy}.
For the neutral-current-mediated decay we find the integrated form~\cite{Gorbunov:2007ak,Bondarenko:2018ptm,Coloma:2020lgy}
\begin{align}
\begin{split}\label{eq:G_nuqq_integrated}
    &\Gamma_{N\to \nu Q\bar Q}= \frac{m_N^5N_CG_F^2U_\alpha^2}{192\pi^3}\Bigg[ \\
    &\, C_1^Q\bigg((1 - 14x^2-2x^4-12x^6)\sigma_x- 4x^4(1-x^4)L_x\bigg)\\
    &\, + 8C_2^Q\bigg((x^2+5x^4-6x^6)\sigma_x - x^4(1-2x^2+2x^4)L_x\bigg)\Bigg],\\
    &\text{ where }L_x = 12\log\left(\frac{2x}{1+\sigma_x}\right)\text{, and } \sigma_x=\sqrt{1-4x^2}. 
\end{split}
\end{align} 
% This is to be compared with the usual definition~\cite{Gorbunov:2007ak,Bondarenko:2018ptm,Coloma:2020lgy}
% \begin{equation}
%     \tilde{L}_x = 3\log \left( \dfrac{1 - 3 x^2 - (1 - x^2)\sigma_x}{x^2
% (1 + \sigma_x)} \right)
% \end{equation}
% which is an equivalent formulation.

\section{The transition point}
\label{sec:transition}

The hadronic and partonic width descriptions discussed above are applicable in their respective energy regimes.
In this section we will first review the state-of-the-art treatment of this transition in Sec.~\ref{sec:naive_mass_dependence}, before attempting to identify a more transparent way of matching these two pictures in Sec.~\ref{sec:differential_ratios}.

\subsection{Mass-dependent branching fractions}\label{sec:naive_mass_dependence}

A common way of estimating the inclusive semileptonic HNL width at a given mass $m_N>1$\,GeV is by evaluating the leading order partonic widths.
The ratio 
\begin{equation}
    R_\tau = \Gamma(\tau\to\nu_\tau+\mathrm{hadrons})/\Gamma(\tau\to\nu_\tau e^-\bar\nu_e)
\end{equation}
can then be recoined as a proxy to relate partonic leading order width with its QCD corrected value, accounting for HNL mass dependence through the running of $\alpha_S$ as a function of $(m_N)$.
A great advantage of $R_\tau$ is that the hadronic width is estimated from $\alpha_S$ at the $\tau$ (or in this case HNL) mass, so that a perturbative expansion is possible (for a recent review on this find e.g. Ref.~\cite{Deur:2023dzc}).
The exact expression for the enhancement is frequently based on the $\mathrm{N^{3}LO}$ in QCD calculation for $R_\tau$ found in Ref.~\cite{Gorishnii:1990vf} yielding 
\begin{equation}\label{eq:parton_enhancement_old}
    \tilde k^\mathrm{N^{3}LO}_\mathrm{LO}=1 +\frac{\alpha_S}{\pi}+ 5.2023\left(\frac{\alpha_S}{\pi}\right)^2 + 26.366\left(\frac{\alpha_S}{\pi}\right)^3.
\end{equation}
In this subsection, we will employ the `contour-improved' $R_\tau$ derived at $\mathrm{N^4LO}$ in $\alpha_S$ including EW corrections~\cite{Baikov:2008jh}, resulting in the expression
\begin{equation}
\begin{split}\label{eq:parton_enhancement}
            k^\mathrm{N^4LO}_\mathrm{LO} = 1.0198\left(\vphantom{\left(\frac{\alpha_S}{\pi}\right)^2}\right.&0.9946  \vphantom{\left(\frac{\alpha_S}{\pi}\right)^5}
                + 1.364\frac{\alpha_S}{\pi} + 2.54\left(\frac{\alpha_S}{\pi}\right)^2\\&\left. + 9.71\left(\frac{\alpha_S}{\pi}\right)^3  + 64.29\left(\frac{\alpha_S}{\pi}\right)^4 \right)\,,
\end{split}
\end{equation}
allowing us to estimate the light-quark partonic HNL width near the $\tau$ mass based on the widths calculated in Sec.~\ref{sec:quark_width}.
For differences in underlying assumptions of the $k$ factor calculation find Ref.~\cite{Baikov:2008jh}.
We approximate $\alpha_S$ using the $\beta$ evolution for the running of the strong coupling~\cite{Deur:2023dzc} up to $\beta_3$, resulting in a $k(m_\tau)-1=0.209$ in good agreement with the experimental value of 0.218~\cite{Baikov:2008jh}.

\begin{figure}
    \centering
    \includegraphics[width=\linewidth]{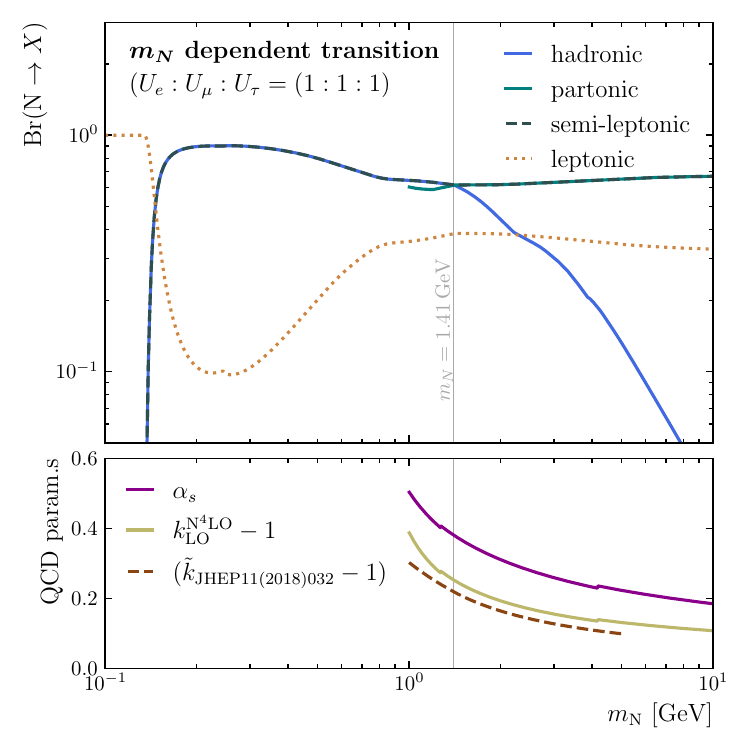}
    \caption{State-of-the-art partial semi-leptonic width estimation.
    Upper: Enhanced partonic (blue), summed hadronic (green), and summed leptonic (dotted orange) branching ratios of an HNL with equal couplings to all lepton generations.
    The semi-leptonic branching fraction (dashed grey) is evaluated from the maximum between partonic and hadronic partial widths. 
    The total width is the sum of semi-leptonic and leptonic partial widths.
    Bottom: Strong coupling $\alpha_S$ using the $\beta$ evolution for the running~\cite{Deur:2023dzc} up to $\beta_3$ (purple), as well as reduced $k^\mathrm{N^4LO}_\mathrm{LO}$ factor between $\mathrm{N^4LO}$ and $\mathrm{LO}$ in QCD of the $\tau$ lepton's partonic width (dark yellow).
    We also show the commonly used $\mathrm{N^3LO}$ as taken from Ref.~\cite{Bondarenko:2018ptm} (dashed brown).}
    \label{fig:mN_transition_point}
\end{figure}

The comparison of summed partial hadronic widths to the enhanced partonic width is shown in the upper panel of Fig.~\ref{fig:mN_transition_point} for the case of an HNL coupling equally to all 3 lepton generations, together with the partonic width enhancement parameters discussed in the text around Eq.(\ref{eq:parton_enhancement}) in the lower panel.
We observe that by including the relevant resonant contributions to the HNL width, we reach a crossover point for HNL masses at around 1.41\,GeV, where the enhanced partonic width grows larger than the summed hadronic states.
This crossover point is sometimes treated in such a way that for larger HNL masses, the difference between the sum of modelled  hadronic and the total partonic decay width is due to higher multiplicity hadronic final states~\cite{Coloma:2020lgy, Feng:2024zfe}.

A notable exception to the common approach outlined above was taken recently by Kretz and Nierste~\cite{Kretz:2025pfk}, where the CC-mediated HNL width was calculated in a fully inclusive way up to $\mathrm{N^4LO}$ from known correlator structures.
In this work they accounted for finite lepton mass effects, not yet including finite quark mass effects in their calculations.
In a stable regime for HNL masses above $\sim2\,\mathrm{GeV}+m_\ell$ (where $\ell$ is the decay product of the CC HNL decay), the close to scale-independent $\mathrm{N^4LO}$ to LO ratio was found to be between 1.15 and 1.17.
We will include this value as an upper benchmark for QCD corrections beyond leading order below.

\subsection{Mass independent hadronisation rates}\label{sec:differential_ratios}
% In the limit of small quark masses ($x_{d}\to0, x_u\to0$ in Eq.s(\ref{eq:G_lud}) and (\ref{eq:G_nuqq})), we find for the differential widths (is this really needed?)
% \begin{align}
% \begin{split}
%     \frac{\mathrm d\Gamma_{N\to \ell U\bar D}}{\mathrm dx} &= \frac{m_N^3N_C|V_{UD}|^2G_F^2U_\alpha^2}{96\pi^3}\sqrt{\lambda(1,x,x_\ell^2)}\\& \times \left((1 - x_\ell^2)^2 + x(1-2x+x_\ell^2)) \right)
% \end{split}\\
% \begin{split}
%     \frac{\mathrm d\Gamma_{N\to \ell U\bar U}}{\mathrm dx}&+\frac{\mathrm d\Gamma_{N\to \ell D\bar D}}{\mathrm dx}=\frac{m_N^3N_C G_F^2U_\alpha^2}{192\pi^3}\\&\times\left(1-2s_w^2+\frac{20}{9}s_W^4\right)(1-x)^2(1+2x)\,.
% \end{split}
% \end{align}

A different way of looking at the two descriptions is revealed when presenting the hadronic-to-partonic ratio of the differential decay widths as a function of the  invariant mass of the hadronic system $q_H$.
Conceptually, this spectral projection is thus equivalent to the hadronisation rates for quark pairs in semileptonic decays for a given scale $q_H$. 
Interestingly, this measure is close to HNL mass independent for HNLs with masses far below that of the $W$ boson, which suggests that the hadronisation ratios only depend on the scale set by $q_H$, not $m_N$ (an example of residual HNL mass dependence can be found in Fig.~\ref{fig:hadronic_differential_decay_widths_mNdep}).
This matches the intuitive expectation that $q_H$ is the scale relating the hadronic system to the QCD vacuum and any interaction between the hadronising system and the $N$--$\ell/\nu$ system is necessarily at least of next-to-leading electroweak order.

\begin{figure}
    \centering
    \includegraphics[width=\linewidth]{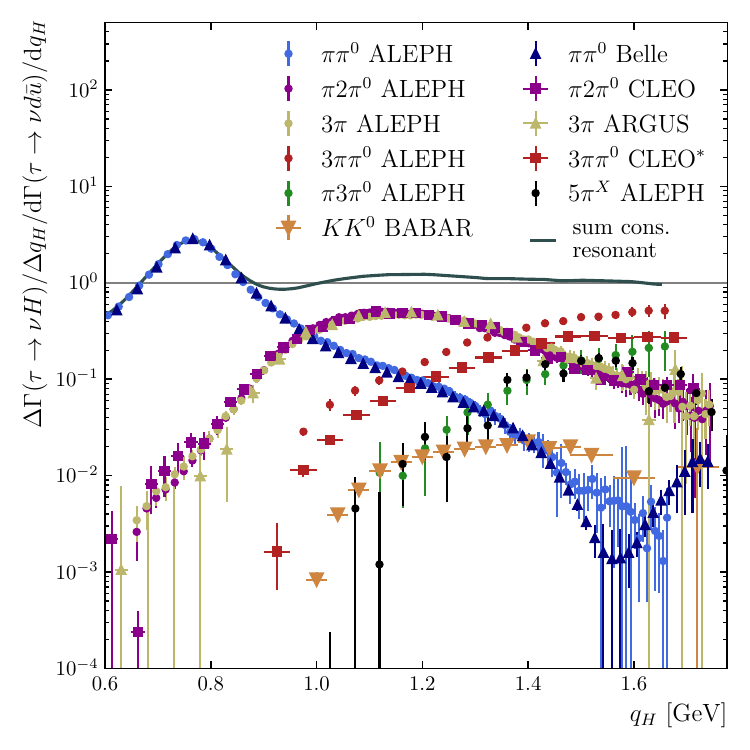}
    \caption{Tau differential partial decay width measurements by the ALEPH collaboration~\cite{ALEPH:2005qgp}, complemented by equivalent measurements by ARGUS~\cite{ARGUS:1992olh}, Belle~\cite{Belle:2008xpe},  CLEO ~\cite{CLEO:1999rzk, CLEO:1999heg}, and BABAR~\cite{BaBar:2018qry} normalised to the $\tau\to\nu \bar u d$ differential cross section given equivalently by Eq.(\ref{eq:G_lud}) for $m_\ell=0$.
    The widths are normalised to the integrated partial widths presented in Table~\ref{tab:tau_decays}, and $9.7\times10^{-3}\Gamma_\tau$ for the summed 5 pion state line~\cite{PDG2026}.
    In the $3\pi\pi^0$ CLEO$^*$ data, the $\pi\omega$ component is subtracted.
    The sum of resonant states is based on ALEPH data.}
    \label{fig:tau_hadronisation_ratio}
\end{figure}

We can, therefore, equivalently present $\tau$ lepton decays in this $\mathrm d\Gamma$-ratio $q_H$ spectral projection.
A corresponding plot of $\tau$ lepton measurements with channels associated with a $u-\bar{d}$ transition is presented in Fig.~\ref{fig:tau_hadronisation_ratio}.
We normalise the partial width measurements to their integral value taken to be the PDG averages listed in Table~\ref{tab:tau_decays}~\cite{PDG2026}.
Clear resonant enhancement, where the hadronic exceeds the partonic differential width, can be observed for the $\rho$ associated peak around 0.78\,GeV, and for the $a_1$ associated peak at around 1.2\,GeV.
We note that large multiplicity final states featuring four or more pions become dominant at around $q_H\geq1.4$\,GeV, where the ratio of hadronic and partonic width becomes close to 1.
In this plot, we list the summed $5\pi$ differential widths measured by ALEPH~\cite{ALEPH:2005qgp}, which are expected to be dominated by the 5 charged pion component ($0.78$ out of $ 0.97$~\% branching ratio).
Up until the kinematic endpoint, their contribution remains sub-leading to that of the $4\pi$ final states.

\begin{figure}
    \centering
    \includegraphics[width=\linewidth]{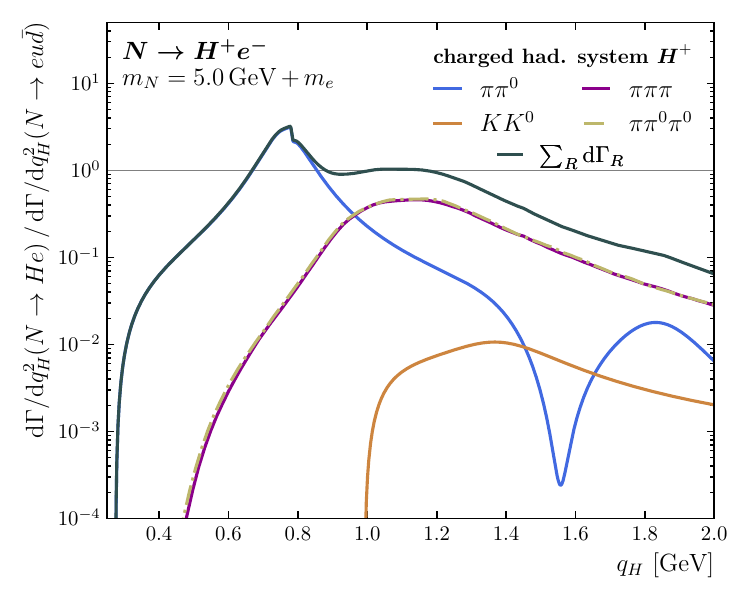}
    \includegraphics[width=\linewidth]{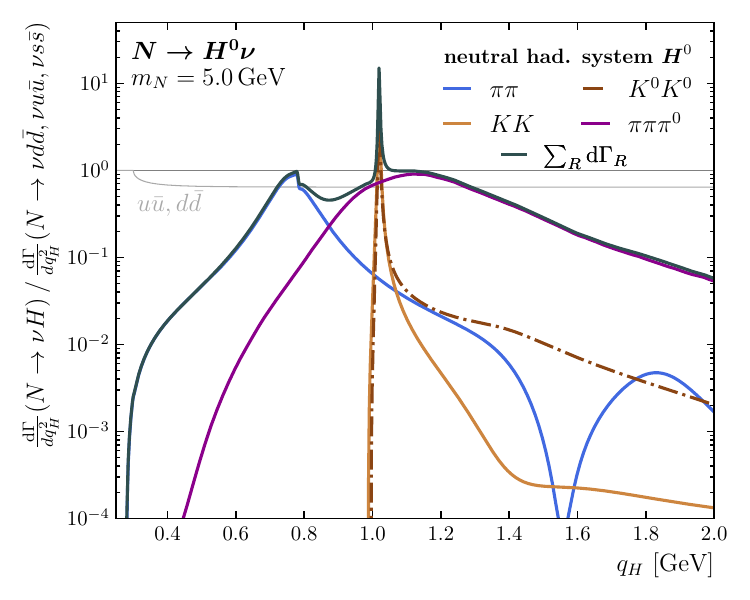}
    \caption{Examples of a hadronic system invariant mass $q_H$ dependent differential width ratio in the charged current decay into an electron (top) and neutral current into a neutrino (bottom). 
    The results are presented for a maximal $q^\mathrm{max}_H=m_N-m_{\ell/\nu}=5$\,GeV, but are independent of HNL and lepton mass.
    The sum of all considered resonant decays is given in dark green.
    For the neutral current, in addition to the $1$ normalisation line in light grey, we show the sum of the $\nu(u\bar u, d\bar d)$ differential widths, indicating the proper normalisation for the $\pi\pi$ and $\pi\pi\pi^0$ states.}
    \label{fig:hadronic_differential_decay_widths}
\end{figure}

Figure~\ref{fig:hadronic_differential_decay_widths} shows such a ratio for an HNL of mass $5\,\mathrm{GeV}+m_\ell$ in charged and neutral current-mediated decays ($m_\ell$ is the lepton mass in CC exchanges and 0 in the NC case). %\footnote{The mass is chosen to guarantee good kinematic coverage of the $q_H$ region of interest. The addition of $m_\ell$ is to assert the same $q_H$ binning for all lepton families, as $q_H\in(\sum_h m_h, m_N-m_\ell)$.}
It includes multi-body hadronic widths as modelled  by the formulas presented in Sect.~\ref{sec:hadronic_decays}, and $u,\,d$ (and $s$) quark widths as given in Eq.s~(\ref{eq:G_lud}) and (\ref{eq:G_nuqq}). 

In the CC ratio, we observe that at hadronic invariant mass below $\sim 1$\,GeV, where the hadronic description is expected to perform well, the partonic width is actually smaller than the partial hadronic HNL width for $q_H$ below $\sim0.7$\,GeV and only becomes larger around the $\rho$ peak at $\sim0.78$\,GeV.\footnote{This does not necessarily signify a complete breakdown of the analogy, as in this regime hadronisation occurs preferably into single meson states.
However, it does reflect the fact that for $q_H$ around and below the hadron scale ($\sim0.8$ GeV), the partonic description is not viable.}
From around 0.9\,GeV to 1.2\,GeV the ratio of summed resonant hadronic widths considered in this work to the partonic widths is close to 1. 
For $q_H$ above 1.2\,GeV, the sum of listed resonant partonic states begins to fall off.
This is in line with expectation from $\tau$ decay measurements, where at 1.25\,GeV the $4\pi$ states begin to surpass 10\,\% of the summed ratios, becoming the dominant decay channel above 1.4\,GeV.
Indeed, for $q_H$ around this value, also \textsc{Pythia} predicts a strong $4\pi$ contribution (see Fig.~\ref{fig:pythia_hadronisation_rates}).
In the resonant picture, these hadronic structures are expected to stem from (broad) $\omega\pi$ and $a_1\pi$ distributions~\cite{CLEO:1999heg}.
From $\tau$ measurements, it is also evident that the $5\pi$ final states (especially the 5 charged pion states) become increasingly relevant for $q_H>1.4$\,GeV.
Other contributions are expected to be of similarly sub-lead\-ing order as the $KK^0$ partial width based on $\tau$ decay branching ratios (see Table~\ref{tab:tau_decays}).
This is especially true as the next biggest contributions in the $\tau$ case come from single kaon final states. 
These are associated with production in the $u\bar{s}$ transition which is not considered here.\footnote{The corresponding modelled  ratios for $u\bar s$ transitions are presented in Fig.~\ref{fig:differential_ratio_usbar}, where we observe a strong resonant peak around in hadronic width for $q_H$ around the $K^\star(892)$ mass.
As no higher multiplicity final states are modelled  in this for this transition, the summed resonant contributions quickly fall off with growing $q_H$, similar to the $\rho$ mediated $2\pi$ final state in the $u\bar d$ transition.}
Given the nature of these stabilizing high multiplicity structures ($\geq4\pi$ states), a transition point in description from hadronic to partonic suggests itself in the range of about $q_H^\mathrm{trans}=1$ to $1.5$\,GeV (see Fig.s~\ref{fig:tau_hadronisation_ratio} and~\ref{fig:hadronic_differential_decay_widths}).
For the remainder of this work, we will set the mean $q_H^\mathrm{trans}=1.25$\,GeV for the light quark channels.
This value might be adjusted if $4\pi$ final states were to be modelled  with clear resonant structure (e.g. through $\rho(1690)$) in future work.

The NC ratios show a similar pattern to the CC case.
The resonant $\rho^0\to\pi^+\pi^-$ and $\phi\to K\bar K$ distributions peak at $q_H\simeq2m_q+0.75\,$GeV, with differential widths exceeding those of the partonic states.
Between $1\,\mathrm{GeV}<q_H\lesssim 1.25$\,GeV the $u\bar u + d\bar d $ partial width is saturated by the $\pi\pi\pi^0$ final state.
For larger $q_H$ we expect higher multiplicity states to saturate the partial quark widths. 
\textsc{Pythia} hadronisation rates suggest significant fractions of multi-pion and di-kaon + multi-pion final states (see e.g. Fig.~\ref{fig:pythia_hadronisation_rates}).
However, the NC $K\bar K$ case is also instructive in highlighting limits of the hadronisation approach. 
While the peak structure for $q_H$ just above $1$ GeV is dominated by a isoscalar transition current  $s\gamma_\mu \bar s$, for larger $q_H$ we observe the effect of non-trivial interference between isoscalar and isovector ($u\gamma_\mu \bar u- d\gamma_\mu\bar d$) contributions, as described in~\ref{sec:kaon_ff}.
The result of this interference for the different partial widths of $K^+K^-$ and $K^0\bar K^0$ final states is shown in Fig.~\ref{fig:hadronic_differential_decay_widths}, which would be difficult to reflect in a hadronisation model.
Nonetheless, the strongly peaking structure also suggests that the total width contribution associated with the $\phi$ meson can be evaluated as a two-body decay in very good approximation.
This also extends to the other basic massive quark vector resonances $D^*$, $D^*_s$, $J/\Psi$, $B^*$, $B_c^*$, and $\Upsilon$ because i) $\phi$ is the broadest of these resonances in absolute terms~\cite{PDG2026}, and ii) analogous interference effects to the one described above for the $\phi$ would likely be mitigated by the distance of interfering resonance structures on the invariant mass spectrum.

\section{Impact on mass-dependent widths}

The above observations can be consistently translated into HNL branching ratios as follows:\\
Heavy neutral lepton decays into broad resonances (e.g. $\rho$, $a_1$, ...) should be treated as full multi-body decays, yielding a good description for hadronic invariant masses $q_H$ below around 1.4\,GeV (see Fig.~\ref{fig:hadronic_differential_decay_widths}).
For matching the resonant hadronic to the partonic picture, two avenues suggest themselves:
\begin{enumerate}[i]
    \item  Evaluating the difference between hadronic and partonic width at a given $q_H$ and using hadronisation fractions to establish the remainder of the hadronic content;
    \item An arguably more elegant solution would be to define a transition point $q_H^\mathrm{trans}$, below which only the hadronic and above which only the partonic picture is applied.
\end{enumerate}
The latter option guarantees that no double counting is applied and is easily extended to also include heavier quark contributions of different mass scales (for this work, this mostly regards constellations featuring one or two charm quarks), by requiring 
\begin{equation}
    q^\mathrm{trans}_{ Q}=q^\mathrm{trans}_{H} + \sum_{{Q_f}} m_{Q_f},
\end{equation}
where $Q_f$ are the final state quarks of a given channel, or the two heaviest quarks constituting a meson.
In both cases, two-body decays of the HNL into metastable mesons (e.g. $\pi$, $K$, $D$, $D_s$, ...) and narrow resonances ($\pi^0$, $\eta$, $\eta^\prime$, $\omega$, $\phi$, $D^*$, ...) can be treated separately, as they are resonantly enhanced in a very narrow region of $q_H$ (of order of resonance width for a comparably small lepton width), so that their partial widths are additive to the established remainder.
This is especially true in ii), where the resonance mass is typically below any sensible choices of $q^\mathrm{trans}_{Q}$, allowing partonic evaluation only above the thresholds of the low lying narrow resonances and metastable mesons. 

As we fully resonantly model only the $u\bar u$, $u\bar d$, and $d\bar d$ channels up to the respective $q^\mathrm{trans}_{ Q}$, we approximate the ha\-dro\-nic widths of the massive quark channels with the lowest lying pseudo-scalar and vector mesons in two-body decays.
We further set $q^\mathrm{trans}_{H}=1.0$\,GeV for these channels, as discussed in~\ref{sec:massive_quark_channels}.

\begin{figure}
    \centering
    \includegraphics[width=\linewidth]{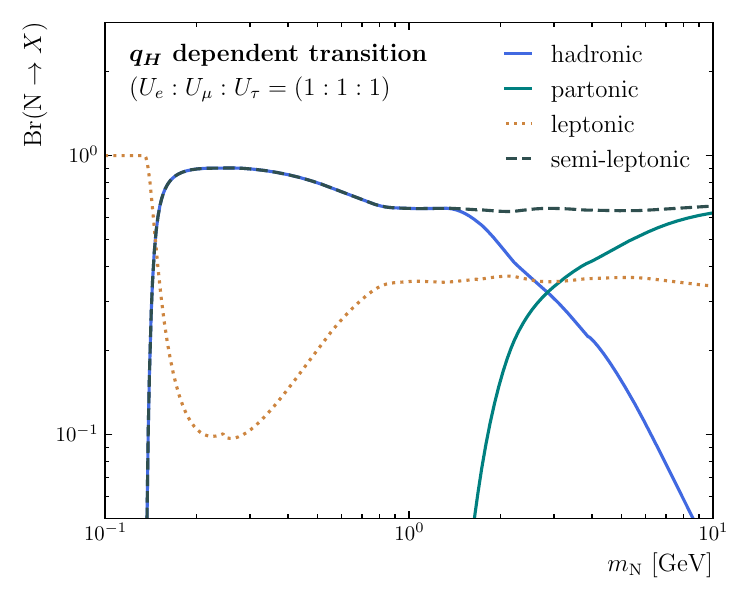}
    \caption{Partonic (blue), summed hadronic (green), and summed leptonic (dotted orange) partial widths of an HNL with equal couplings to all lepton generations, assuming a $q^\mathrm{trans}_{H}=1.25$\,GeV.
    The semi-leptonic partial width (dashed grey) is considered as the sum of partonic and hadronic, and the total width is the sum of semi-leptonic and leptonic partial widths.
    To be contrasted against Fig.~\ref{fig:mN_transition_point}.}
    \label{fig:regime_split_branching_mdep}
\end{figure}

For the truncated quark widths, the $k$ factor in Eq.(\ref{eq:parton_enhancement}) is no longer applicable, as it encapsulates also effects at low $q_H$.
Therefore, we apply only the tree-level expressions for quark widths. 
The resulting correction should be less significant than corrections to $R_\tau$, as the widths are only evaluated in a regime where pQCD becomes applicable.
We show the resulting HNL partial widths to hadronic, partonic, and leptonic final states in Fig.~\ref{fig:regime_split_branching_mdep} for an HNL with equal couplings to all generations as a key point of this study. It  should be contrasted against Fig.~\ref{fig:mN_transition_point}.
In this prescription, semi-leptonic final states are now the sum of hadronic and partonic partial widths.
We find that for HNL masses below 1.7\,GeV the HNL semi-leptonic decay width is almost fully (>90\,\%) described in the hadronic picture.
% This is just slightly below the transition point of 1.90\,GeV established in the naive mass-dependent picture in Sect.~\ref{sec:naive_mass_dependence}.
For HNL masses between 2 and 6 GeV, hadronic and partonic descriptions both remain relevant, while for even larger masses ($m_N>7.6$\,GeV) the semi-leptonic width could well be approximated (>90\,\%) in the partonic picture.
If one were to disregard any transition to the partonic picture, the total HNL width for an HNL with mass of the common production threshold of the strange $D$-meson mass ($m_N=m_{D_s}$) would be underestimated by 8\,\%.

\begin{figure}
    \centering
    \includegraphics[width=\linewidth]{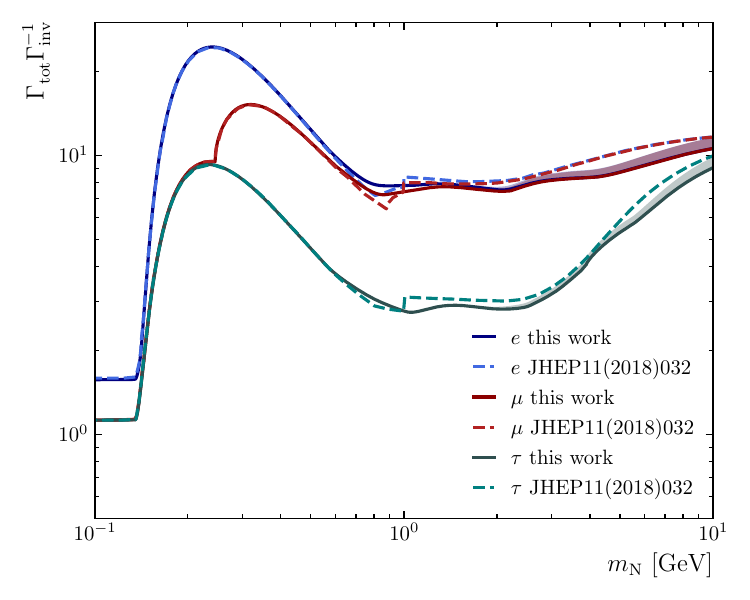}
    \caption{HNL mass-dependent total width normalised to the $3\nu$ partial width for an electrophilic (blue), muonphilic (red), and tauphilic (green) HNL, assuming a $q^\mathrm{trans}_{H}=1.25$\,GeV.
    We also include a comparison to the equivalent quantities given by Bondarenko et. al \cite{Bondarenko:2018ptm} indicated in dashed lines with the same colour scheme. 
    The shaded areas above the solid width lines correspond to a flat multiplicative factor of 15\,\% to the parton contributions as indicated might be applicable from beyond leading order HNL studies~\cite{Kretz:2025pfk}.}
    \label{fig:norm_total_branching_ratios_mdep}
\end{figure}

In Fig.~\ref{fig:norm_total_branching_ratios_mdep} we show the resulting total HNL decay widths for HNLs coupling only to either electrons, muons, or taus normalised to the `invisible' width
\begin{equation}\label{eq:G_inv}
    \Gamma_\mathrm{inv}=\sum_{\beta}\Gamma\left(N\to\nu_\alpha\nu_\beta\bar\nu_\beta\right) = U^2_\alpha \frac{G_F m_N^5}{192\pi^3}.
\end{equation}
We also include a comparison to the commonly used prescription established by Bondarenko et al.~\cite{Bondarenko:2018ptm} and observe notable consequences of the treatment discussed in this work.
For HNL mass below 1\,GeV, the width differences are main\-ly due to the $3\pi$ and $KK$ final states that were not considered in Ref.~\cite{Bondarenko:2018ptm}. 
Up to HNL masses of around 1.8 GeV, we expect our hadronic width estimate to largely reflect the full HNL width, as hadronic tau decay channels (and their NC counterparts) not accounted for in this work sum to $\tau$ branching fractions below 5\%.
The discrepancy for 1\,GeV$<m_N<1.8$\,GeV with Ref.~\cite{Bondarenko:2018ptm} indicates that the $k$-factor width estimation employed there is prone to overestimating the semileptonic width, especially when applying the $R_\tau$ derived for CC decays to NC channels.
To mitigate this effect, other recent works on the subject resort to ad-hoc kinematic corrections of the $s\bar s$ width~\cite{Coloma:2020lgy,Feng:2024zfe}.

For 2\,GeV$<m_N$, partonic effects begin to become impactful (see also Fig.~\ref{fig:regime_split_branching_mdep}).
We indicate a 15\,\% increase of the total partonic width as evaluated in this work by a the shaded areas in Fig.~\ref{fig:norm_total_branching_ratios_mdep} as is indicated to be a stable $\mathrm{N^4LO}$ to LO ratio in Ref.~\cite{Kretz:2025pfk} for massless quarks.
This number is likely an overestimation of beyond LO effects, as in the case of Ref.~\cite{Kretz:2025pfk}, it also accounts for strongly resonantly enhanced effects.
However, even including this enhancement, our width estimates remain smaller than those given by Ref. \cite{Bondarenko:2018ptm}.
We believe that this is mainly due to our including heavy quark contributions only above the respective vector meson, rather than quark mass thresholds.
The remaining difference at very large masses ($m_N\to10$\,GeV) is mostly due to not including the abovementioned $k$ factor enhancement, as in this limit also our prescription is completely dominated by the partonically established width.

\begin{figure}
    \centering
    \includegraphics[width=\linewidth]{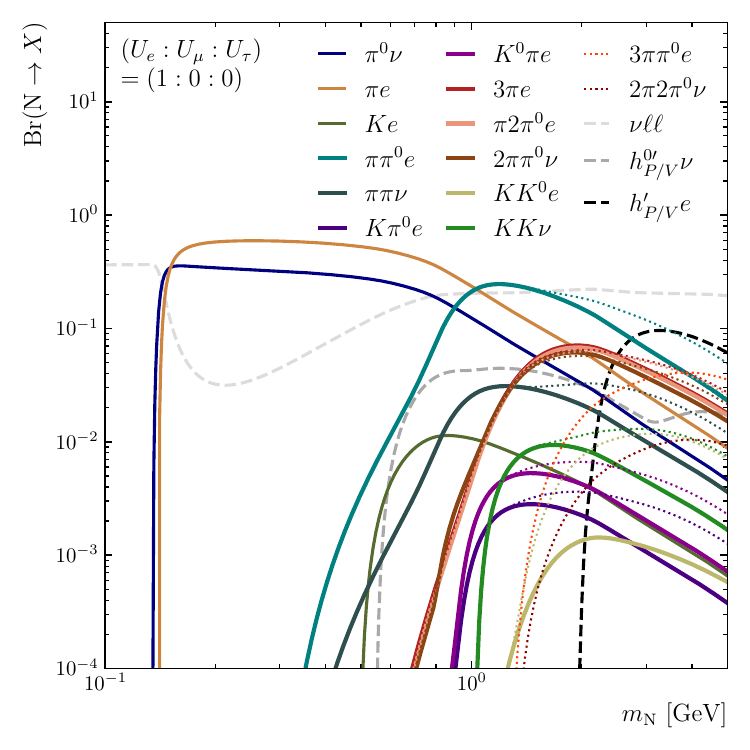}
    \caption{Branching ratios of an electrophilic HNL as calculated under the prescription of this work.
    The underlying total width is calculated as outlined in the text and can be taken from Fig.\ref{fig:norm_total_branching_ratios_mdep}.
    Solid lines indicate partial widths calculated purely in the resonant approximation (ie. $q_H^\mathrm{trans}\to\infty$).
    Dotted lines indicate partial widths derived from \textsc{Pythia} with $q_H^\mathrm{trans}=1.25$\,GeV ($q_Q^\mathrm{trans}=1.$,GeV for $u\bar s$ states $K^0\pi$, $K\pi^0$).
    The dashed lines indicate sums of other decay classes: light grey (dark grey / black)  is the sum over three-body charged leptonic states (residual neutral two-body width /charged two-body widths) dominated by $\nu e e$ ($\nu\eta^{(\prime)}$ / $D_s e$) at low masses and $\nu e \mu$ ($J/\Psi \nu$ / $D_s^* e$) at larger mass values.
    }
    \label{fig:branching_mN_U2el}
\end{figure}

To translate the partonic widths (for $q_H>q_Q^\mathrm{trans}$) into well-defined hadronic signatures of the HNL decay a hadronisation tool has to be applied.
To demonstrate the idea, we use the hadronisation rates given by \textsc{Pythia}8.3~\cite{Bierlich:2022pfr} as described in~\ref{sec:pythia_hadronisation}, folded with the applicable HNL leading-order differential widths to quark pairs as given by the integrands in Eq.s (\ref{eq:G_lud}) and (\ref{eq:G_nuqq}), to derive mass-dependent branching ratios.
We caution that this approach does not lead to satisfactory results and is only used to illustrate the concept.
The resulting branching ratios for an electrophilic HNL under the prescription of this work are presented in Fig.~\ref{fig:branching_mN_U2el}.
We show the hadronic branching ratios evaluated without $q_H$ cut on hadronic evaluation as solid lines.
The dashed dark grey (black) line indicates the sum over the parts of decays into neutral two-body pseudo-scalar and vector resonances presented in \ref{sec:hadronic_two_body_pars} that are not included as effective channels towards the solid final states.
The light grey dashed curve is sum of branching ratios from of the well-established~\cite{Gorbunov:2007ak,Bondarenko:2018ptm, Coloma:2020lgy} two lepton widths for $\nu ee$, $\nu e \mu$, and $\nu\mu\mu$.\footnote{Their widths are related to the partonic widths presented in Sect.~\ref{sec:quark_width}, where $\Gamma(N\to\nu_\alpha \ell_\alpha \bar\ell_\beta)$ is given by Eq.~(\ref{eq:CC_3_body}), and $\Gamma(N\to\nu_\alpha \ell_\beta \bar\ell_\beta)$ by Eq.(\ref{eq:G_nuqq_integrated}), with ($C_1^f= (1+(2\delta_{\alpha\beta}-1)4s_W^2+8s_W^4)/4$, $C_2^f= s_W^2(2s_W^2+2\delta_{\alpha\beta}-1)/2$)~\cite{Gorbunov:2007ak,Helo:2010cw,Atre:2009rg,Bondarenko:2018ptm,Coloma:2020lgy}. }
The dotted lines, emerging from solid lines representing multi-body states, correspond to the branching ratios if an evaluation via partonic final states hadronised by \textsc{Pythia} is considered.

\section{Impact on experimental sensitivities}
\label{sec:sensitivities}

\begin{figure}
    \centering
    \includegraphics[width=\linewidth]{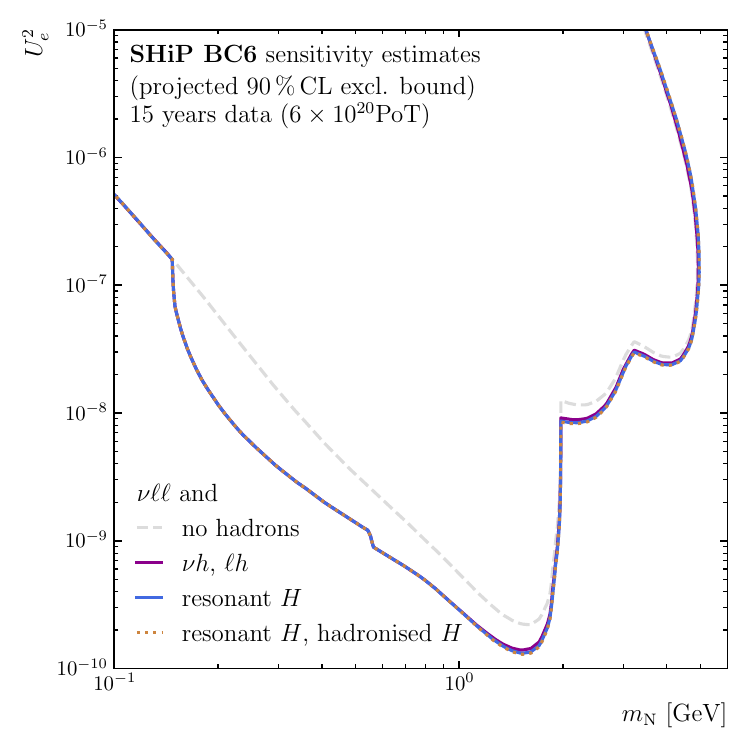}
    \caption{Estimated sensitivity of the planned SHiP experiment estimated with \textsc{Alpinist} based $D$ and $B$ meson spectra generated with \textsc{Pythia}~8.3, assuming $\sigma_{cc}=20\,\mu$b and $\sigma_{bb}=2.63\,$nb~\cite{Schubert:2024hpm}.
    Presented are several assumptions on the hadronic HNL final states as presented in this work in addition to the leptonic final states 
    $\nu\ell\ell$.
    For details on the hadronic assumptions see text.
    }
    \label{fig:SHiP sensitivity}
\end{figure}

There are numerous projects under consideration or in preparation that experimentally target unexplored HNL parameter space (see, e.g. Fig 8.19 of Ref.~\cite{deBlas:2025gyz}).
A few of these experiments that have recently actively targeted HNLs in parameter space where the findings of this work are relevant include CMS, LHCb, ATLAS, and NA62~\cite{ CMS:2024ake, LHCb:2025ymr, ATLAS:2025uah, Schubert:2026puf}.
To outline the effects of the prescription described in this work, we present the expected sensitivity of the planned SHiP experiment at CERN's ECN3 experimental facility after a full 15 years of data taking.~\cite{SHiP:2025ows}
We evaluate the sensitivity in \textsc{Alpinist}, a toy-MC simulation framework, enabling us to study the sensitivity impact of several hardon theory assumptions based on largely theory-independent acceptance estimates for defined final states.
The SHiP experiment is suitable for such a comparison, as it aims for sensitivity covering the full MeV to GeV HNL mass spectrum.
It also satisfies the required kinematic assumption of simulations in \textsc{Alpinist} that production and decay of the are sufficiently far apart.\footnote{We would like to highlight that \textsc{Alpinist} is capable of estimating sensitivities for any combination of HNL coupling ratios $(U^2_e:U^2_\mu:U^2_\tau)$, as was presented in Ref.~\cite{Schubert:2024hpm}.
% We emphasize as well that a number of other proton beam dumps such as NA62, NuCal, CHARM, SEAQUEST, and DUNE-ND are implemented in \textsc{Alpinist}~\cite{Jerhot:2022chi, Schubert:2024hpm}, so the current study can be straightforwardly extended by running the public software framework from Ref.~\cite{Alpinist_Zenodo}.
}
The geometry and cuts on final states are taken from the SHiP technical proposal~\cite{SHiP2023} and summarised in~\ref{sec:SHiP_cuts}.

The resulting sensitivity estimates are presented for an electrophilic HNL in Fig.~\ref{fig:SHiP sensitivity} produced in charmed and beauty meson decays (excluding $B_c$)~\cite{Schubert:2024hpm}. 
In this sensitivity estimate, we restrict ourselves to the well-defined final states $\nu ee$, $\nu e\mu$, $\nu \mu\mu$, $\pi^0\nu$, $\pi\pi\nu$, $\pi e$, $\pi\pi^0e$ $2\pi\pi^0\nu$, $3\pi e$, $\pi2\pi^0 e$, $2\pi2\pi^0\nu$, and $3\pi\pi^0 e$ with branching ratios as presented in Fig.~\ref{fig:branching_mN_U2el}.
The sensitivity is presented as a projected 90\,\%CL exclusion limit under zero-background hypothesis with 100\,\% reconstruction efficiency.
As a baseline, we show the sensitivity projection for only leptonic final states (grey dashed), which is the only source of sensitivity for HNL masses below the pion mass. 

In purple, we show the projected sensitivity for hadronic final states in the two-body decay approximation ($\nu h$ and $\ell h$) as was implemented in previous versions of \textsc{Alpinist}~\cite{Schubert:2024hpm}, modelled  after Ref.~\cite{Bondarenko:2018ptm}, but with updated parameters as presented in~\ref{sec:decay_pars}.
We observe that over the full kinematic range where they are available, hadronic final states are major drivers of experimental sensitivity, highlighting the importance of their accurate description.
The resonant contributions up to $q_H^\mathrm{trans}=1.25\,$GeV presented in this work are shown in blue, where we consider only the resonantly produced final states.
From around $m_N=$1\,GeV up to the kinematic limit at 5.3\,GeV, we observe a marginal sensitivity improvement over the two-body approximation, mostly driven by the $3\pi$ states at the bottom edge of the sensitivity scaling as the square root of the number of visible decays with a flat decay profile along the decay volume.
A slight increase in sensitivity is also observed at the top part of the sensitivity curve, driven by the HNL lifetime, reflecting how many HNLs decay before reaching the fiducial volume.
Here, the increase is due to the longer lifetime associated with a smaller total width calculated in the prescription outlined in this work (see Fig.~\ref{fig:norm_total_branching_ratios_mdep}).

Finally, in dotted orange, we present the sum of resonant and hadronised contributions, as evaluated based on \textsc{Pythia}~8.3 hadronisation rates (as presented in ~\ref{sec:pythia_hadronisation}).
A very modest sensitivity increase in the regime around $m_N=2$\,GeV is observed, mainly associated with the in\-creas\-ed $\pi\pi^0 e$ branching fraction due to the $\pi\pi^{(0)}$ hadronisation ratio given by \textsc{Pythia}.
For masses of $m_N\simeq5$\,GeV the sensitivity estimate is almost exclusively given by the leptonic final states, even when including the dominant five-body final states $3\pi\pi^0e$ and $2\pi2\pi^0\nu$ as taken from \textsc{Pythia}.
This is due to the extremely large ratio of high multiplicity final states, including 5 or more hadrons which cannot reasonably be encapsulated in a framework of searches for specific final state signatures.
In consequence, the partonic width becomes depleted in terms of `observable' hadronic final states.
This is not the case for the well defined leptonic $\nu\ell\ell$, which ends up being the most relevant contribution close to the kinematic limit at around 5.3\,GeV, even though their branching fraction remains close to constant for $m_N$ between 1 and 5 GeV (see Fig.\ref{fig:branching_mN_U2el}).
We stress again that, due to the poor matching of 2-body hadronic final states, this should not be regarded as a final estimate but serves only to highlight the significance of hadronised states per se.
Under the assumption that the multiplicity estimates produced in this naive hadronisation estimate are representative, hadronic states in the partonic regime (especially for $q_H>2.5$\,GeV) are likely better estimated by searches for jet-like structures.

For all simulated final states, we assumed a flat phase space profile of the decay products.
This is justified as even for a complicated decay structure like the $\pi2\pi^0e$ decay, effects due to a differential phase space modelling are negligible (see~\ref{sec:phase_space_modelling}).

Looking beyond the sensitivity impact of a broader range of final states, a potential difference in estimated total width (and thus HNL lifetime) is important in and of it self.
This is particularly so, for displaced vertex searches as performed at ATLAS, CMS and LHCb, where the HNL lifetime strongly impacts the selection efficiency of the displacement matching algorithms~\cite{LHCb:2025ymr,ATLAS:2025uah,CMS:2024hik}.
However, a dedicated investigation of these effects goes beyond the scope of this paper.

\section{Conclusions}
\label{sec:conclusions}
    
In this work, we investigated semi-leptonic decays of heavy neutral leptons with masses in the GeV regime.
For the first time, we explicitly included the three pion decays mediated by the $a_1$ resonance, as well as full kinematic treatment of three-body final states involving kaon-kaon and kaon-pion pairs. 
We also investigated possible avenues of consistently linking the resulting hadronic to the partonic decay picture, focusing on a transition based on the invariant mass of the hadronic system $q_H$.

Based on the differential ratios derived in Sect.~\ref{sec:differential_ratios}, we propose the following treatment for hadronic decays of HNLs in the GeV regime:
\begin{enumerate}
    \item HNL decays into broad resonances (e.g. $\rho$, $a_1$, ...) should be treated as full multi-body decays up to hadronic invariant masses $q^\mathrm{trans}_H\simeq$1.2 to 1.5\,GeV;
    \item For invariant masses greater than $q^\mathrm{trans}_H$, hadronisation fractions for the given quark pairs are folded with the relevant differential quark-level partial widths;
    \item HNL decays into into into metastable mesons (e.g. $\pi$, $K$, $D$, $D_s$, ...) and narrow resonances ($\eta$, $\eta^\prime$, $\omega$, $\phi$, ...) can be treated as additive to the partial widths, due to the strong resonant enhancement in a $q$ regime where the partonic mechanism is not evaluated.
\end{enumerate}
We find that for HNL masses below $2\,$GeV, the resonant description fully covers the relevant phenomenology.
Regardless of whether or not this approach is employed for estimating the total HNL width, the effective low energy description, as for example outlined in this work, should be applied for simulating final states with hadronic invariant masses below the perturbative regime ($q_H\lesssim q_Q^\mathrm{trans}$), irrespective of the HNL mass.

In terms of experimental sensitivity, we found that considering purely resonant $3\pi$ final states has limited impact with respect to considering only $\pi$ and $2\pi$ decays, for zero-background experiments such as NA62 in dump mode or the forthcoming SHiP experiment.
For HNL masses above 2\,GeV hadronised partonic contributions become significant.
Naive estimates based on \textsc{Pythia} hadronisation ratios suggest that even including the dominant 3 and 4 $\pi$ final states as well-defined hadronic final states does not lead to a significant sensitivity improvement in this mass range.
This is because even higher multiplicity final states become increasingly common, `depleting' the partonic width in terms of experimental signatures defined in the hadronic picture.

In this work we preliminarily used an untuned version of \textsc{Pythia}~8.3 to estimate the hadronisation fractions with a transition hadronic invariant mass of $q_H^\mathrm{trans}=1.25$\,GeV.
Achieving a better match between the differential partonic and the established hadronic widths established in this work could possibly be done by tuning standard hadronisation tools such as \textsc{Herwig}~\cite{Bellm:2025pcw} or \textsc{Pythia}~\cite{Bierlich:2022pfr} to fully match the partial width ratios at $q_H^\mathrm{trans}$.
Alternatively, one could opt for a data-driven approach such as the \textsc{ScalarHadroni\-zer} for Higgs-like scalars~\cite{Gieseke:2025gfq}.
Indeed, this approach is already implemented in \textsc{Alpinist} for Higgs-like scalars~\cite{Dobrich:2026mvi}.
However, as the underlying conserved quantities differ in semileptonic HNL decay, this implementation is not directly applicable and has to be reimplemented. 
To facilitate the development of a better matching hadroniser, other resonant HNL decays into kaons~\cite{Finkemeier:1995sr} or final states involving 4 pions along the lines of Ref.~\cite{Czyz:2008kw}, or by including $\rho(1690)$ and other resonances, could be adapted.
A shortcoming of such an approach, as of now, is the lack of differential phase space distributions for the hadronised particles. 
For zero-background experiments, we expect this to play only a sub-leading role (see e.g. \ref{sec:phase_space_modelling}).
The resulting experimental impact for searches with defined hadronic final states for $m_N>2$\,GeV will have to be revisited in light of such a dedicated hadronisation tool, especially regarding the abundance of high multiplicity final states in the associated $q_H$ regimes between 2 and 5 GeV.
We aim to investigate the hadronisation of the partonic system more closely in a future publication.

Another main consequence of the treatment put forward in this work is the impact on the total HNL decay width shown in Fig.~\ref{fig:norm_total_branching_ratios_mdep}.
The decreased total width and associated increase of HNL lifetimes for a given point in $m_N$-$U^2_\alpha$ parameter space would likely lead to an increase in sensitivity for displaced vertex searches at collider experiments~\cite{CMS:2024hik,ATLAS:2025uah,LHCb:2025ymr} at the upper mass limit, determined largely by the HNL lifetime.
However, we caution that the widths were provided at LO in the partonic picture, with contributions beyond leading order likely not negligible.
A first step toward going beyond LO was recently taken in Ref.~\cite{Kretz:2025pfk}, relating fully inclusive $\mathrm{N^4LO}$ CC-mediated HNL decays to the $\mathrm{LO}$ results, with a resulting width increase of order 1.15 for relevant HNL masses and scales.
We included this ratio as a flat uncertainty estimate in the total width presented in Fig.~\ref{fig:norm_total_branching_ratios_mdep}.
For the approach suggested in this work, a more straightforward application of pQCD at the scale $q_H$ is applicable.
Here, the NLO corrections to off-shell polarised gauge boson decays into massive quarks are known~\cite{Groote:2013xt} and likely applicable.
We defer the investigation of QCD corrections of the partonic widths to future work.
\begin{acknowledgements}

We thank Stefan Wallner for sharing his expertise regarding the $a_1$ resonance, 
as well as Jan Jerhot, Felix Kahlh{\"o}fer, and Tommaso Spadaro for many fruitful discussions regarding this work.
The authors gratefully acknowledge funding through the European Research Council under grant ERC-2018-StG-802836 (AxScale) as well as through the Deutsche For\-schungs\-ge\-mein\-schaft (DFG, German Research Foundation) under Germany’s Excellence Strategy – EXC 2094 – 39078331.
\end{acknowledgements}
\appendix
\section{Mass dependence of hadronisation ratios}

\begin{figure}
    \centering
    \includegraphics[width=\linewidth]{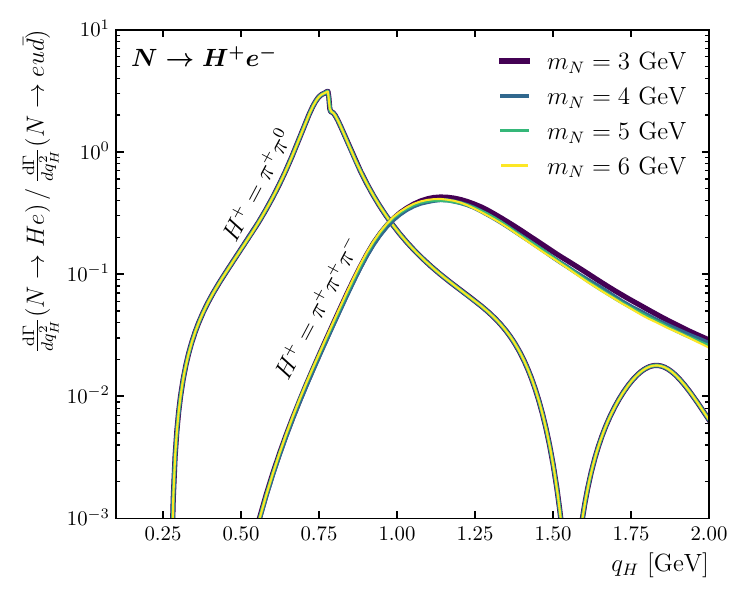}
    \caption{Hadron system mass dependence of the differential width ratios in $N\to e\pi\pi^0$ and $N\to e\pi\pi\pi$  decays presented for different HNL mass values. 
    In the $\pi^+\pi^0$ case, differences between ratios for different HNL masses are below plot resolution.}
    \label{fig:hadronic_differential_decay_widths_mNdep}
\end{figure}

The mass dependence of the differential width ratio with respect to the hadronic mass is shown in Fig.~\ref{fig:hadronic_differential_decay_widths_mNdep} for the $2\pi$ and $3\pi$ systems in a charged current mediated decay of an electrophilic HNL.
We note thatthe differential width ratio of the $2\pi$ decay is indistinguishable between the presented masses at the presented resolution.
The differential width ratio for the $3\pi$ decay shows minimal variation between different masses, especially as $q_H$ becomes comparable to $m_N$.

\section{Sensitivity dependence on phase-space modelling}\label{sec:phase_space_modelling}

\begin{figure}
    \centering
    \includegraphics[width=\linewidth]{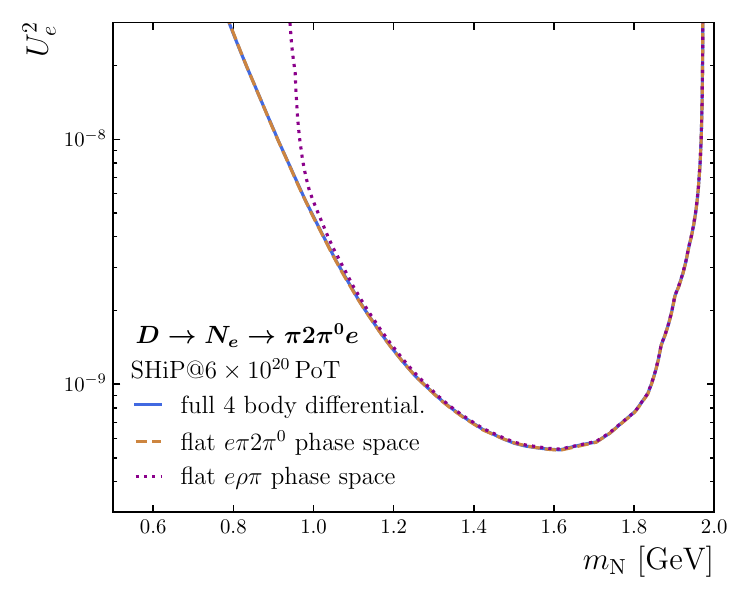}
    \caption{90\,\%CL sensitivity of the SHiP experiment to an electrophilically produced HNL decaying into $\pi2\pi^0 e$ assuming accurate 4-body phase space (red), flat 4-body phase space (blue) or flat 3-body phase space with an on-shell intermediate rho meson (orange).}
\label{fig:ship_d_pi2pie_sens}
\end{figure}

In Fig.~\ref{fig:ship_d_pi2pie_sens}, we show the single channel 90\,\%CL sensitivity of the SHiP experiment to an electrophilically produced HNL decaying into $\pi2\pi^0 e$ under varying assumptions on the decay phase space as evaluated using \textsc{Alpinist} using cuts and detector model presented in Ref.~\cite{SHiP2023}. 
We note that for zero-background counting experiments such as NA62~\cite{Schubert:2026puf} or the planned SHiP experiment~\cite{SHiP:2025ows}, the differential structure of the phase space has negligible impact.
For experiments such as CMS or LHCb, where signal and background are to be distinguished on a differential level, this picture might be significantly altered. 
With this work, we provide an event list for different HNL masses of resonant $a_1$, mediated $N\to3\pi\ell$ and $N\to\pi2\pi^0\ell$ for $\ell\in(e,\mu,\tau)$, available within \textsc{Alpinist}\cite{Alpinist_Zenodo}.

\section{Hadronisation of virtual vector bosons}
\label{sec:pythia_hadronisation}

One possible avenue of estimating the abundance of hadronic final states in partonic HNL decays is to hadronise the quark pair using a standard hadronising tool and counting the number of occurrences of stable final states with respect to the number of events.
The results of such an approach for $Z\to u\bar u $ and $W\to u\bar d $ are shown in Fig.~\ref{fig:pythia_hadronisation_rates}.
Here, we initialise a vector boson of virtuality $q_H$ ($E=q_H$, $\vec{p}=0$) in \textsc{Pythia}~8.3 and allow only decays into the indicated quark pairs.

\begin{figure}
    \centering
    \includegraphics[width=\linewidth]{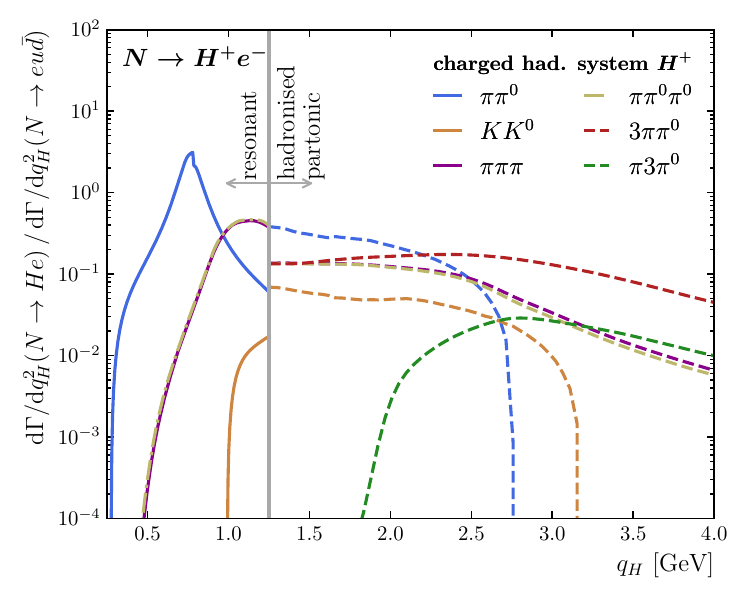}
    \includegraphics[width=\linewidth]{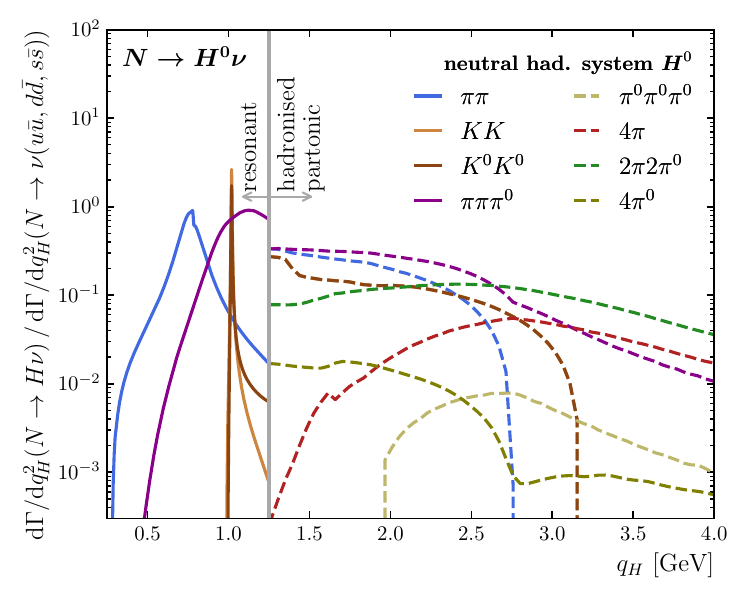}
    \caption{Hadron partial width ratio calculated in the resonant picture ($q_H<1.25\,$GeV, solid lines) and in the hadronised partonic picture ($q_H>1.25\,$GeV, dashed lines) as taken from \textsc{Pythia}8.3 and weighted by the respective quark widths.
    This is shown for charged (top) and neutral current (bottom) mediated decays.}
\label{fig:pythia_hadronisation_rates_transition}
\end{figure}

\begin{figure}
    \centering
    \includegraphics[width=\linewidth]{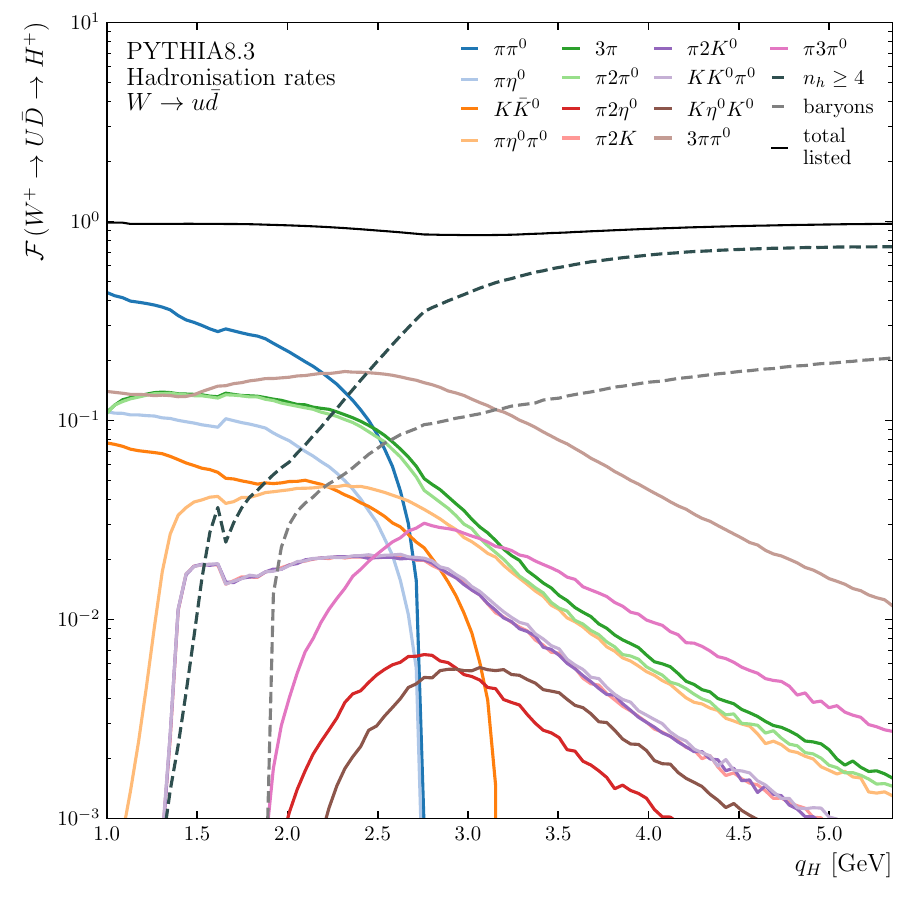}
    \includegraphics[width=\linewidth]{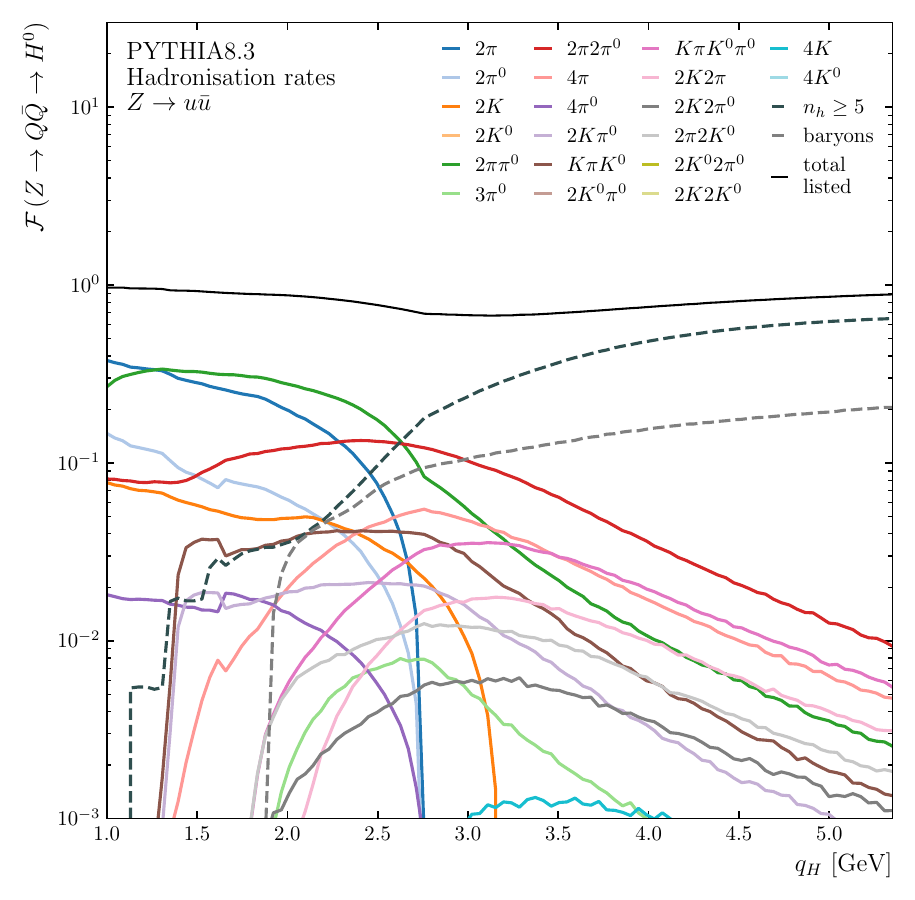}
    \caption{Hadronisation ratios for off-shell vector-bosons decaying into quark pairs (top: $W^+\to u\bar d$, bottom: $Z^0\to u\bar u$) as given by \textsc{Pythia}~8.3.
    Here `baryons' denotes any final states including protons or neutrons, and $n_h$ is the multiplicity of purely mesonic final states not listed in the legend.
    The remainder of final states not accounted for is due to final states involving photons and leptons.
    Results are reported as ratios of observed events in 200 thousand simulations per mass bin.}
\label{fig:pythia_hadronisation_rates}
\end{figure}

\textsc{Pythia} could be tuned to match the hadronisation ratios established in this work.
Here, we limit ourselves to presenting the hadronisation ratios for an untuned version of \textsc{Pythia} for a transition scale $q^\mathrm{trans}=1.25\,$GeV in Fig.~\ref{fig:pythia_hadronisation_rates_transition}.
This scale was chosen, as the resonant contributions listed in this work fall below the threshold of the anticipated partonic width  above this value and have to be compensated by decays into not-considered higher multiplicity states (e.g. $4\pi$) based on $\tau$ decay structure (see Fig.~\ref{fig:tau_hadronisation_ratio}).
We note thateven an untuned version of \textsc{Pythia} performs somewhat well in estimating the hadronisation ratio for the 3 pion final states at this scale, especially in the neutral current case in line with expectations from $\tau$ data (Fig.~\ref{fig:tau_hadronisation_ratio}).
However, by enforcing the decays as described above, \textsc{Pythia} overestimates the relative abundance of especially the states with even numbers of hadrons.
In both the neutral and charged current cases, the importance of the 2 pion and 2 kaon final states are overestimated by \textsc{Pythia} with respect to the resonant calculation.
What's more, the \textsc{Pythia} result also includes a strong $2\pi^0$ component in the neutral decay channel, which is incompatible with stemming from a rho-resonance-mediated hadronisation, see upper panel of Fig.~\ref{fig:pythia_hadronisation_rates}).

Notably, \textsc{Pythia} predicts a sizeable fraction of the quark pair hadronisation to result in final states featuring baryons.
Another key observation is thathigh multiplicity final states quickly dominate, making search approaches traditionally targeted at specific final states (as for example performed at LHCb~\cite{LHCb:2025ymr} or NA62~\cite{Schubert:2026puf}) impractical.

\section{Decay parametrisation} \label{sec:decay_pars}

In this part, we briefly present the parametrisations employed to evaluate the semi-leptonic decays considered in this work.

\subsection{Two body decay constants} \label{sec:hadronic_two_body_pars}

The hadronic currents underlying the calculations of Sect.~\ref{sec:two_body_decays} are
\begin{align}\label{eq:hadronic_current_definitions}
    &J^\mu_{h_P^+} =if_hp^\mu, &J^\mu_{h_P^0} =-i\frac{f_h}{\sqrt{2}}p^\mu, \\
    &J^\mu_{h_V^+} =ig_h\epsilon^\mu(p), &J^\mu_{h_V^0} = i\frac{\kappa_h g_h}{\sqrt{2}}\epsilon^\mu(p),
\end{align}
where $p^\mu$ is the meson's momentum, $f_h$ are the pseudo-scalar decay constants in Tables~\ref{tab:charged_decay_constants} and~\ref{tab:neutral_decay_constants}, $\epsilon$ is the vector meson's polarisation vector, and $g_h$ its decay constant with the dimensionless constant $\kappa_h$ relating charged and neutral meson decays. 
The charged and neutral vector meson decay constants are given in Tables~\ref{tab:neutral_vector_decay_constants} and~\ref{tab:charged_vector_decay_constants} .
These normalisations were chosen to keep consistency with previous literature on the subject~\cite{Bondarenko:2018ptm, Coloma:2020lgy}.
For comparing the decay constants reported in Table~\ref{tab:neutral_decay_constants} to literature values outside of the works on HNLs, it is important to note that our definition of $f_\eta$ differs from that commonly used in reporting $f^\mathrm{lit}_\eta$ in literature as $f_\eta= f^\mathrm{lit}_\eta/\sqrt{2}$, to match the underlying normalisations in Eq.(\ref{eq:hadronic_current_definitions}).
Similarly, for vector mesons, a common definition in literature for vector decay constants is $f_{h_V} = g_h/m_h$, where different conventions exist regarding a factor $\sqrt{2}$.

The decay constants of charged pseudo-scalars $\pi$, $K$, $D$, and $D_s$ can be calculated in lattice QCD. 
Recent summary results are presented in Table~\ref{tab:charged_decay_constants}.

\begin{table}[ht]
    \centering
    \begin{tabular}{c|cccccc}
        \hline
        $h^\pm_P$ & $\pi$ & $K$ & $D$ & $D_s$ & $B$ & $B_c$\\
        \hline
        $f_h$ [MeV] & 130.2 & 155.7  & 210.4  & 247.7 & 192.0 &  378 \\
        \hline
    \end{tabular}
    \caption{Leptonic decay constants $f_h$ of charged pseudo-scalar mesons $h^\pm_P$ from lattice QCD.\cite{FLAG:2024oxs,Tao:2022qxa}}
    \label{tab:charged_decay_constants}
\end{table}

The decay constants for neutral pseudo-scalars can then be related/approximated as follows.
From $\chi\mathrm{PT}$ in the limit of exact isospin, we get 
\begin{equation}
    f_{\pi^0}=\sqrt{2}F^0=f_{\pi^\pm},
\end{equation}
where $F^0$ is the $\chi\mathrm{PT}$ pseudo-Goldstone boson decay constant normalised in $\mathrm{Tr}(\lambda^a\lambda^b)=2\delta^{ab}$ convention (here $\lambda^a$ are the Gell-Mann matrices).

Although the mixing of $\eta$ and $\eta^\prime$ makes a well-defined leptonic decay constant for these mesons technically impossible~\cite{Feldmann:1999uf}, we employ recent lattice results~\cite{Ottnad:2025zxq} in Feldmann-Kroll-Stech (FKS) scheme~\cite{Feldmann:1998vh} to derive approximate values.\footnote{The decay constants enter only as sub-leading contributions to the hadronic width estimates.
Even for $\tau$-philic HNLs where the NC contributions greatly outweigh the CC, due to the large lepton mass, the maximum branching fraction $\mathrm{Br}^{\max}_{N\to\nu\eta}<7$\,\% which occurs around $m_N\simeq1$\,GeV.}
In the FKS scheme, the axial vector structure in the quark flavour basis is taken to be
\begin{align}\label{eq:FKS_axial_currents}
    A^\mu_l =& \frac{1}{\sqrt{2}}(\bar u \gamma^\mu\gamma_5 u + \bar d \gamma^\mu\gamma_5 d)\,\text{, and}\\
    A^\mu_s =&\bar s \gamma^\mu\gamma_5 s\,, 
\end{align}
where the decay constants relate to those of the $\eta$ and $\eta^\prime$ mesons by a single mixing angle
\begin{equation}
    \left(\begin{array}{cc}
f_l^\eta & f_s^\eta \\
f_l^{\eta^{\prime}} & f_s^{\eta^{\prime}}
\end{array}\right)=\left(\begin{array}{cc}
f_l \cos \phi & -f_s \sin \phi \\
f_l \sin \phi & f_s \cos \phi
\end{array}\right) .
\end{equation}
In this scheme the decay constant parameters $f_l$ and $f_s$ do not exhibit scale-dependent contributions.
We then take the literature values ($f_l=138.6(4.4)$\,MeV, $f_s=170.7(3.3)$\,MeV,  $\phi=0.686(35)$) from lattice QCD~\cite{Ottnad:2025zxq} and relate the $\eta^{(\prime)}$ quark structure to the currents in Eq.(\ref{eq:FKS_axial_currents}), thus deriving
\begin{align}
    f^\mathrm{eff}_\eta &\simeq \frac{f_l^{\eta}- \sqrt{2}f_s^{\eta}}{\sqrt{6}}=106.2\,\text{MeV, and} \\
    f^\mathrm{eff}_{\eta^\prime} &\simeq \frac{\sqrt{2}f_l^{\eta^\prime}+f_s^{\eta^\prime}}{\sqrt{6}}= 104.6\,\text{MeV,}
\end{align}
where we accounted for the $\sqrt{2}$ difference in normalisation.
These results are of similar absolute value as found in Ref.s~\cite{Bondarenko:2018ptm, Coloma:2020lgy} and in good agreement with those quoted in Ref.~\cite{Atre:2009rg}, when accounting for the $\sqrt{2}$ normalisation.
For the $\eta_c$ and $\eta_b$, well-defined decay constants can be computed using lattice QCD.\cite{Davies:2010ip,Hatton:2021dvg}

\begin{table}[ht]
    \centering
    \begin{tabular}{c|ccccc}
        \hline
        $h_P^0$ & $\pi^0$ & $\eta$ & $\eta^\prime$ & $\eta_c$ & $\eta_b$ \\
        \hline
        $f^\mathrm{(eff)}_h$ [MeV] & 130.2 & 106.2& 104.6 & 279.1 & 511.9 \\
        \hline
    \end{tabular}
    \caption{Leptonic decay constants $f_h$ of neutral pseudo-scalar mesons $h_P^0$ derived from lattice QCD results.
    See discussion in text.}
    \label{tab:neutral_decay_constants}
\end{table}

The neutral current correction factors $\kappa_h$ for neutral vector mesons can be inferred in the meson-mediated picture of the neutral current~\cite{Schubert:2024hpm, Coloma:2020lgy}.
The associated decay constants are then derived from the electromagnetically mediated decays of these mesons into leptons, which are well measured~\cite{PDG2026} by relating 
\begin{equation}
    g_h = \frac{m_h }{c_\mathrm{EM}\alpha_\mathrm{EM}}\sqrt{\frac{3m_h}{4\pi}\Gamma(h\to e^+e^-)}, 
\end{equation}
where $\alpha_{EM}$ is the electromagnetic coupling constant, and $c_\mathrm{EM}$ is the neutral meson's constituent quark's electric charge.
The resulting factors and decay constants are presented in Table~\ref{tab:neutral_vector_decay_constants}.
In this work only the neutral vector mesons $\omega$, $\phi$, $J/\Psi$, and $\Upsilon$ are considered. 
To the derive the $\Upsilon$ coupling value, we use $\alpha_\mathrm{EM}=1/132.15$ as evaluated at the bottom mass.~\cite{Pivovarov:2000cr}

\begin{table}[ht]
    \centering
    \begin{tabular}{ccc}
        \hline
         $h^0_V$ & $\kappa_{h}$ & $g_h$ [GeV$^2$] \\
         \hline
         $\omega$   & $-\frac{2}{3}\sin^2\theta_W$                           & 0.157 \\
         $\phi$     & $\frac{2\sqrt 2}{3}\sin^2\theta_W - \frac{1}{\sqrt 2}$ & 0.232 \\
         $J/\Psi$   & $\frac{1}{\sqrt 2} - \frac{4\sqrt 2}{3}\sin^2\theta_W$ & 1.29  \\
         $\Upsilon$ & $\frac{2\sqrt 2}{3}\sin^2\theta_W - \frac{1}{\sqrt 2}$ & 6.42  \\
         \hline
    \end{tabular}
    \caption{Neutral vector current correction factors $\kappa_{h}$ and vector meson decay constants $g_h$~\cite{Coloma:2020lgy, Schubert:2024hpm} for neutral vector mesons $h^0_V$.}
    \label{tab:neutral_vector_decay_constants}
\end{table}

\begin{table}[ht]
    \centering
    \begin{tabular}{c|cccc}
        \hline
        $h^\pm_V$ & $D^*$ & $D_s^*$ & $B^*$	& $B_c^*$ \\
        $g_V$ [GeV$^2$] & 0.470 & 0.579 & 0.932 & 2.22\\
         \hline
    \end{tabular}
    \caption{Decay constants $g_V$ of charged vector mesons $h^\pm_V$ from Lattice QCD~\cite{Dhiman:2017urn, Chen:2020qma, Tao:2022qxa}.}
    \label{tab:charged_vector_decay_constants}
\end{table}

\subsection{Three-body vector decay parametrisations}
\label{sec:three_body_vector_current}
\label{sec:kaon_ff}

% \subsection{Kaon vector form factor}
For a pair of pseudo-scalar mesons $h\bar h$, the axial currents of the form $\langle h|q \gamma^\mu\gamma_5\bar q|h\rangle$ vanish due to parity.
The charged current then takes the purely isovector form 
\begin{equation}
    j_\mu^\mathrm{CC} = \bar u \gamma_\mu d +\bar u\gamma_\mu s.
\end{equation}
The vector form factor is then defined through 
\begin{equation}
    \langle h^0h^-| \bar d \gamma^\mu u |0\rangle = \left(p_{h^0}-p_{h^-}\right) F^h_{V} (s).
\end{equation}
with eventual strange quarks as spectators in the transition.

\begin{figure}
    \centering
    \includegraphics[width=\linewidth]{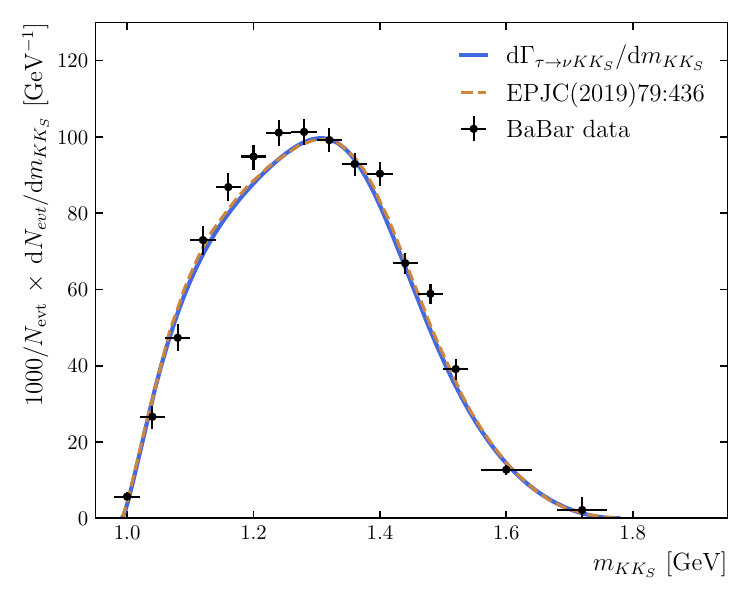}
    \caption{Normalised number of $\tau\to \nu KK_s$ events as a function of $KK_s$ invariant mass observed by BaBar~\cite{BaBar:2018qry} and expected from Eq.(\ref{eq:N_ellhh}) based on $F^K_{V}$ parametrisations. 
    See text for detail.} 
    \label{fig:kaon_fv}
\end{figure}

We model the kaon vector form factor $F^K_{V}$, using a para\-metri\-sation in Omn{\`e}s exponential representation (fit  `iv' based on Eq.(51) in Ref.~\cite{Gonzalez-Solis:2019iod}).\footnote{We find that the presented results disagree strongly with our evaluation under the nominal fit results but find good agreement for $\tilde\phi_1=-1.088$ (instead of the quoted $\tilde\phi_1=-1.88$). 
Consequently, we use $\tilde\phi_1=-1.088$ in the associated figure, as well as the form factor employed in this work. }
A comparison between BaBar $\tau\to \nu K^-K_S$ data~\cite{BaBar:2018qry}, the presented (orange dashed), and our results with adjusted $\tilde\phi_1$ (blue) is shown in Fig.~\ref{fig:kaon_fv} in terms of observed and expected events per $m_{K K_S}$ invariant mass bin of $40\,$MeV width.

The electromagnetic (EM) current for the pseudo-scalar pair decomposes as 
\begin{equation}
    j_\mu^\mathrm{EM} = \tfrac 12\left(\bar u \gamma_\mu u -\bar d\gamma_\mu d\right) +  \tfrac 16\left(\bar u \gamma_\mu u + \bar d\gamma_\mu d\right) -  \tfrac 13\bar s\gamma_\mu s,
\end{equation}
consisting of an isovector part $\tfrac 12(\bar u \gamma_\mu u -\bar d\gamma_\mu d)$ and an isoscalar remainder.
The neutral weak current is
\begin{align}
    &j_\mu^\mathrm{NC} = j_\mu^\mathrm{T3}-2\sin^2\theta_W j_\mu^\mathrm{EM},\,\text{where}\\
    &j_\mu^\mathrm{T3} = \tfrac 12 \left(\bar u \gamma_\mu u -\bar d\gamma_\mu d\right) - \tfrac 12 \bar s\gamma_\mu s.
\end{align}

\begin{figure}
    \centering
    \includegraphics[width=\linewidth]{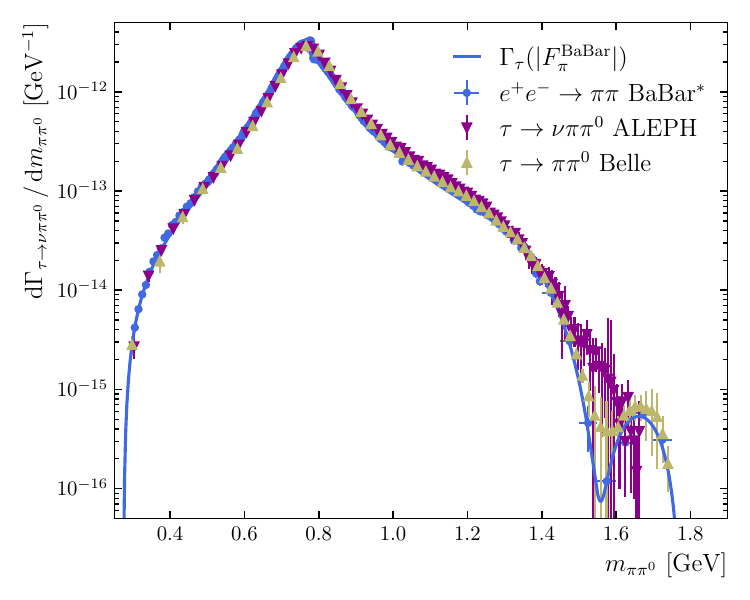}
    \caption{Differential width of $\tau\to\nu\pi\pi^0$ as given by Eq.~(\ref{eq:N_ellhh}) with pion form factor as used in this work. This is overlaid with the $e^+e^-\to\pi\pi$ measurement by BaBar~\cite{BaBar:2012bdw} incorporated into Eq.~(\ref{eq:N_ellhh}) with an underlying $\sigma(s)\propto\beta_3(s)/s|F_\pi|^2$ (blue), as well ALEPH~\cite{ALEPH:2005qgp} (purple) and as Belle~\cite{Belle:2008xpe} (yellow) measurements of $\tau\to\nu\pi\pi^0$ decays.}
    \label{fig:tau_pipi_ff_pi}
\end{figure}

For a pair of pions, the isoscalar transition part $\propto (\bar u \gamma_\mu u + \bar d\gamma_\mu d)$ is forbidden by G-parity, yielding the proportionality $j_\mu^\mathrm{NC}=(1-2\sin^2\theta_W)j_\mu^\mathrm{EM}$.
Further, $\propto (\bar u \gamma_\mu u + \bar d\gamma_\mu d)$ can be related to $u\gamma_\mu\bar d$ by isospin rotation so that the form factors associated with the currents $\sqrt{2}f^{\mathrm{EM}v}_\pi\simeq f^{\mathrm{CC}v}_\pi$.
We follow Ref.~\cite{Bondarenko:2018ptm} in modelling the pion form factor $f^{\mathrm{CC}v}_\pi$ as related to the BaBar fit to $e^+ e^- \to \pi^+\pi^-$ scattering with contributions from $\rho$, $\omega$, $\rho^\prime$, $\rho^{\prime\prime}$, and $\rho^{\prime\prime\prime}$ (given by Eq.(26) of Ref.~\cite{BaBar:2012bdw}).
A comparison with experimental measurements~\cite{ALEPH:2005qgp,Belle:2008xpe} in $\tau$ decays as shown in Fig.\ref{fig:tau_pipi_ff_pi} shows good agreement.
To infer this comparison we use the proportionality of the measured cross section $\sigma$ to the form factor $\sigma(s)\propto\beta_3(s)/s|F^\mathrm{EM}_\pi|^2$, and apply Eq.~(\ref{eq:N_ellhh}) to calculate the corresponding the differential width.
The numerical integrals of all displayed differential widths are normalised to the $\tau\to\nu\pi\pi^0$ partial width PDG value of $5.78\times10^{-13}$\,GeV.

For a pair of kaons, the form factor needs to be split into an isoscalar and an isovector component. 
In the isovector transition, the strange quark acts purely as a spectator.
Thus, like in the pion case, we can relate the isovector part of the NC to that of the CC through isospin rotation, leading to the relationship $f_K^{\mathrm{EM}v}=F_K^{\mathrm{CC}V}/\sqrt{2} = (1-\sin\theta_W^2)f_K^{\mathrm{NC}v}$.

The isoscalar component of the EM case, $f_K^{\mathrm{EM}v}$ has recently been modelled in Ref.~\cite{Stamen:2022uqh}.
Unfortunately, the $s\gamma_\mu \bar s$ component in $j^\mathrm{T3}$ breaks the direct proportionality of EM and NC currents and, thus, their form factors.
Nonetheless, relating $f_K^\mathrm{EMs}$ to the EM current structure, we can again employ the  meson-mediated picture of the current, where 
\begin{equation}
    f_K^{{\mathrm{EM}}s} = \tfrac 16\left(\bar u \gamma_\mu u + \bar d\gamma_\mu d\right) -  \tfrac 13\bar s\gamma_\mu s \equiv \tfrac 16 f_K^\omega - \tfrac 13 f_K^\phi.
\end{equation}
From this we can now assemble $F^\mathrm{NC}=F^\mathrm{T3}-2\sin^2\theta_WF^\mathrm{EM}$, leading to the neutral current isoscalar form factor 
\begin{equation}
    f_K^{{\mathrm{NC}}s}=  \frac{4\sin\theta^2_W-3}{6}f_K^\phi- \frac{\sin\theta^2_W}{3}f_K^\omega,
\end{equation}
so that 
\begin{equation}
    F^\mathrm{NC}_{K^+K^-} = f_K^{{\mathrm{NC}}s} + f_K^{{\mathrm{NC}}v}\,\text{and }\, F^\mathrm{NC}_{K^0\bar K^0} = f_K^{{\mathrm{NC}}s} - f_K^{{\mathrm{NC}}v}.
\end{equation}

% \begin{align}
%     F^\mathrm{NC}_{K^+K^-} &=\left(1-2\sin\theta^2_W\right)f^{{\mathrm{EM}}s} - \left( \frac{3-4\sin\theta^2_W}{6}f_K^\phi+ \frac{\sin\theta^2_W}{3}f_K^\omega\right)           \\
%     -F^\mathrm{NC}_{K^0\bar K^0} &= \left(1-2\sin\theta^2_W\right)f^v+\left( \frac{3-4\sin\theta^2_W}{6}f_K^\phi+\frac{\sin\theta^2_W}{3}f_K^\omega \right) .
% \end{align}
The resonance specific currents in the EM fit are given by~\cite{Stamen:2022uqh}
\begin{align}
    &f_K^\phi = c_\phi \frac{M_\phi^2}{M_\phi^2-s-i\sqrt{s}\Gamma_{\phi}(s)},\text{ and}\\
    &f_K^\omega = c_\omega \frac{M_\omega^2}{M_\omega^2-s-iM_\omega\Gamma_\omega},
\end{align}
and a higher resonance $\omega^\prime$ regulator for the normalisation and large $Q$-limit behaviour 
\begin{equation}
    f^{\omega^\prime}=(3-2 c_\phi - c_\omega)\frac{M_{\omega^\prime}^2}{M_{\omega^\prime}^2-s-iM_{\omega^\prime}\Gamma_{\omega^\prime}}.%\left(3+2 c_\phi - c_\omega - 3\frac{1-c_\phi}{2\sin\theta^2_W} \right)
\end{equation}
The fit values in the EM case are largely compatible with the $\mathrm{SU}(3)$-symmetric limit expectation $c_\phi=c_\omega=1$.\footnote{For the $K^0\bar K^0$ fit these are ($c_\phi=1.001$, $c_\omega=1.07$), and for $K^+K^-$ found to be ($c_\phi=0.977$, $c_\omega=1.42$).}
Lacking a corresponding clean SM channel to fit to, we will adopt $c_\phi=c_\omega=1$ for the NC form factor.
The dynamic $\phi$ width is parametrised as~\cite{Hoferichter:2014vra}
\begin{equation}\begin{aligned}
   &\Gamma_\phi(s) = \sum_{K=K^+,K^0}\frac{\gamma_{\phi\to \bar K K}(s)}{\gamma_{\phi\to \bar K K}(M_\phi^2)}\Gamma^\prime_{\phi\to \bar K K} \theta\big(s-4M_K^2\big) \\ 
   &\quad+ \frac{f_{\phi\to\pi\rho+3\pi}(s)}{f_{\phi\to\pi\rho+3\pi}(M_\phi^2)}\Gamma^\prime_{\phi\to\pi\rho+3\pi} \theta\big(s-(M_\rho+M_\pi)^2\big),
\end{aligned}\end{equation}
where the $\Gamma^\prime$ refer to the partial widths rescaled to compensate for all other decay channels not included explicitly,
\begin{align}
\begin{split}
	\gamma_{\phi\to \bar K K}(s) &= \frac{\left(s-4M_K^2\right)^{3/2}}{s},\text{and}\\
	f_{\phi\to\pi\rho+3\pi}(s) &= \left(\frac{\lambda\left(s,M_\rho^2,M_\pi^2\right)}{s}\right)^{3/2}.
\end{split}
\end{align}
To assure the right interference pattern underlying the modelling in of $F^{\mathrm{EM}s}$, we use an unsubtracted dispersion integral 
\begin{equation}
f_K^{\mathrm{NC}v}(s)=\frac{1-2\sin\theta_W^2}{\sqrt{2}\pi}\int_{4\pi^2}^\infty \mathrm{d} s' \frac{\mathrm{Im} \left(F_K^V(s')\right)}{s'-s} \label{eq:iv-DR}
\end{equation}
to evaluate the vector form factor, as was done in Ref.~\cite{Stamen:2022uqh}.

\begin{figure}
    \centering
    \includegraphics[width=\linewidth]{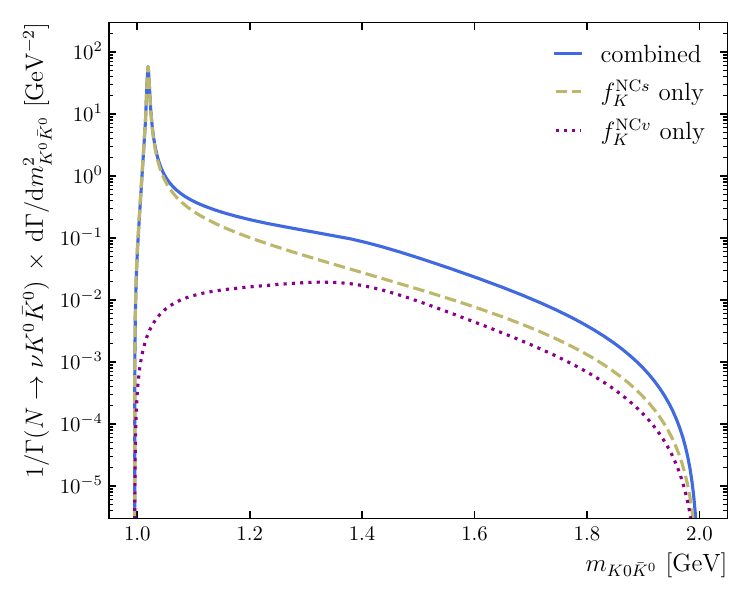}
    \caption{Isoscalar $f_K^{\mathrm{NC}v}$ and isovector $f_K^{\mathrm{NC}v}$ form factor impact on the $N\to K\bar K^0$ differential partial width for an HNL of 2\,GeV mass.
    } 
    \label{fig:kkbar_ff}
\end{figure}

The resulting contributions of the form factors $f_K^{\mathrm{NC}v}$ and $f_K^{\mathrm{NC}v}$ to the $N\to K^0\bar K^0$ differential partial widths for $m_N=2$\,GeV are shown in Fig.~\ref{fig:kkbar_ff}.
The integrated total width is largely driven by the peak below $q_H<1.1$\,GeV (82\,\% of partial width), 97\,\% of which are due only to the isoscalar contribution. 
% \TD{How is this expected to relate to $\nu\phi$ state? Depending on $m_N$ the $\Gamma$ 3/2 body ratio varies from 0.3 near threshold to ... >0.6 ? Comment on how it extends to high-quark-mass equivalents}

For modelling $F^{V/S}_{K\pi}$, we employ a VMD fit by the Belle collaboration to $\tau\to \nu K_s\pi$ data~\cite{Belle:2007goc}.
As the scalar form factor mediated by neutral $K^*$ resonances is mostly responsible for modelling threshold effects, we rely on the model fit focussing on the vector resonances, which also features the smallest reduced $\chi^2$.
This fit incorporates contributions from $K^*_0(700)$, $K^*(892)$ and $K^*(1410)$.
The Finkmeier-Mirkes sca\-lar form factor $F^S_\mathrm{FM}(Q^2)$ employed in the Belle fit is related to the normalisation in Eq.(\ref{eq:2hadron_current}) by
\begin{equation}
    F^S(Q^2) = \frac{m_{K}^2-m_\pi^2}{Q^2} F^S_\mathrm{FM}(Q^2).
\end{equation}
Following Ref.~\cite{Finkemeier:1996dh}, we relate the $K^- \pi^0$ and $\bar K^0 \pi^-$ form factors by a simple Clebsch-Gordan factor
\begin{equation}
    F^{V/S}_{ K^- \pi^0} = \frac{1}{\sqrt{2}}F^{V/S}_{\bar K^0 \pi^-}
\end{equation}
in good agreement with branching ratio measurements yielding a ratio of $\mathrm {Br}(\tau\to\nu K^- \pi^0) = 0.49\mathrm {Br}(\tau\to\nu \bar K^0 \pi^-)$~\cite{PDG2026}.

\begin{figure}
    \centering
    \includegraphics[width=\linewidth]{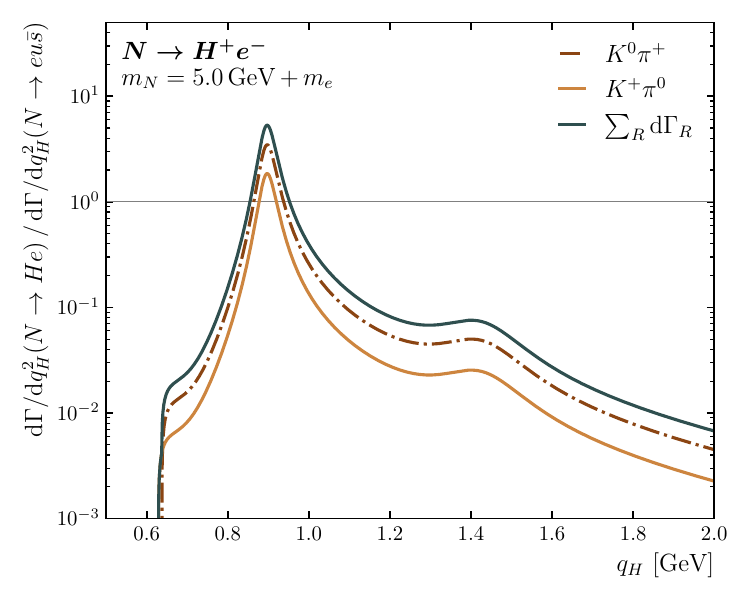}
    \caption{Differential width ratio for $K\pi$ final states produced in $u\bar s$ transitions normalised to the underlying differential quark widths. 
    The hadronic width is modelled after Eq.(\ref{eq:2hadron_current}). 
    For assumptions about the form factors see text. }
    \label{fig:differential_ratio_usbar}
\end{figure}

The resulting differential widths in $q_H$ spectral projection are given as a ratio to the $N\to \nu u\bar s$ differential width in Fig.~\ref{fig:differential_ratio_usbar}.
The $K\pi$ to $u\bar s$ width ratio of the largely reproduces the same $q_H$ dependence pattern as $\pi\pi^0$ to $u\bar d$ as was presented in Fig.~\ref{fig:hadronic_differential_decay_widths}.
We observe a modest increase from threshold to vector resonance mass, with a pronounced peak above unity, which falls off with larger $q_H$.
For larger $q_H>1$ GeV, we expect higher multiplicity states such as $K\pi\pi$ to stabilise the width ratio towards 1.

\subsection{Partial waves of the $a_1$ mediated decay}\label{sec:a1_partial_waves}

The hadronic current of the $a_1\to3\pi$ decays can be  decomposed into partial waves.
We take the pion momenta to be $p^\mu_i$, $P^\mu_{ij}=p_i^\mu +p^\mu_j$, $P^\mu_{123}=P^\mu_{12}+p^\mu_3$, and assume the isobar resonance $R$ to enter as shown in Fig.~\ref{fig:a1_decay_structure}, with $p_1$ being the momentum pion not affected by eventual symmetrisation.
The hadronic currents can be decomposed into several partial wave components, namely an S-wave 
\begin{equation}\label{eq:s_wave}
    J_S^\mu = n_\pm\left(G^{\mu\nu}\left( \mathrm{BW}_R(P_{12}^2)K^\nu_{12} \pm (2\leftrightarrow3)\right)\right),
\end{equation}
a P-wave,
\begin{equation}
    J_P^\mu = G^{\mu\nu} \mathrm{BW}_R(P_{12}^2)\left(P_{12}^\nu -p_3^\nu\right),
\end{equation}
and the D-wave 
\begin{equation}\label{eq:d_wave}
    J_D^\mu = n_\pm\left(\left( K^\mu_{123}K^\nu_{123} -G^{\mu\nu} K^2_{123}\right) \mathrm{BW}_R(P_{12}^2)K^\nu_{12} \pm (2\leftrightarrow3)\right).
\end{equation}
Here, doubled Lorentz-indices are to be contracted, and we have defined the transversal projection tensor
\begin{equation}
    G^{\mu\nu}= g^{\mu\nu} - \frac{P_{123}^\mu P_{123}^\nu}{P_{123}^2},
\end{equation}
with $g^{\mu\nu}$ as the Minkowski metric, the $ij$ transverse momentum 
\begin{equation}
    K^\mu_{ij} = \frac12\left(p_i^\mu -p_j^\mu -\frac{(m_i^2-m_j^2)P_{ij}^\mu}{P_{ij}^2}\right)
\end{equation}
and the full transverse momentum 
\begin{equation}
    K^\mu_{123} = \frac12\left(P_{12}^2 - p_3^\mu -\frac{(P_{12}^2-m_3^2)(P_{123}^\mu)}{P_{123}^2}\right).
\end{equation}
In principle this expansion could be continued to higher order~\cite{Kuhn:1992nz,Krinner:2021let}, but we will limit the discussion in this work to these components.
The Bose-symmetrisation $+(2\leftrightarrow3)$ has to be applied if two identical particles are present in the final state which is the case in the CC-mediated decays involving $\rho$ resonances in the intermediate states. 
As for the NC-mediated decay all final state particles are unique, no such symmetrisation is applied and the two different contributions mediated by $\rho^+$ and $\rho^-$ are summed coherently.
{We find that the sum of amplitudes amounts to $-(2\leftrightarrow3)$ in Eq.s(\ref{eq:s_wave}) and (\ref{eq:d_wave}), as from an isospin perspective $|a_1^0\rangle\propto(|\rho^+\rangle|\pi^-\rangle - |\rho^-\rangle|\pi^+\rangle)$.
The absence of symmetrisation incurs a factor of $\sqrt{2}$ at the current level, or factor 2 at the differential width, so that
\begin{equation}
    n_\pm=\frac12\left((1+\sqrt{2})\pm(1-\sqrt{2})\right).
\end{equation}
This reflects that the neutral $a_1$ width, is equal to that of the charged $a_1$, up to interference effects between $\rho^+$ and $\rho^-$ mediated amplitudes, proportional to $\Gamma_\rho/m_\rho$.
The $P$ wave component for the $a_1$ occurs only for scalar resonances with the identical final state particles in the isobar system, so that again no symmetrisation, but also no further summation is required.

We model $ij$ isobar resonances using Breit-Wigner functions, convention i.e.
\begin{equation}\label{eq:basic_bw}
    BW_h(s) = \frac{m_0^2}{m_0^2-s-im_0\Gamma(s)},
\end{equation}
with dynamic width
\begin{equation}
    \Gamma(s)= \Gamma_0\frac{m_0}{\sqrt{s}} \left(\frac{\gamma(s) }{\gamma(m_0^2)}\right)^{2L+1},
\end{equation}
where $L$ is the orbital momentum between the final states and
\begin{equation}
    \gamma(s) = \sqrt{\frac{\lambda(s,m_i^2,m_j^2)}{4s}}.
\end{equation}

Measurements of the $a_1$ resonance in $\tau$ decays are available in the all-charged-$\pi$ case~\cite{ARGUS:1992olh,ALEPH:2005qgp}, as well as the mostly neutral case~\cite{CLEO:1999rzk,ALEPH:2005qgp} with more recent preliminary results from Belle~\cite{Rabusov:2022woa,Rabusov:2023tna}.
The relevant parameters are listed in Table~\ref{tab:a1_parameters}.
% , where the partial wave amplitudes $\beta_R$ are given for BW propagators in $\mathrm{BW}(q^2)\xrightarrow[]{q^2\to0}1$ convention for comparison to literature, which relate to the amplitudes $A_R$ in this work by 
% \begin{equation}
%     A_R = \frac{m_0}{\Gamma_0} \beta_R,
% \end{equation}
% modulo an overall factor of $i$.
The $m_{a_1}^0$ and $\Gamma_{a_1}^0$ measured by ARGUS are compatible with the PDG averages of $1202\pm15$\, and $422\pm12$\,MeV respectively which include measurements in $\pi p$ scattering and $D^0$ decays, as well yielding a similar $D$ to $S$ wave amplitude ratio (PDG ratio is $-0.062\pm0.020$)~\cite{PDG2026}.
CLEO model the $a_1$ propagator including an alternative effective $a_1$ mass running with a dispersive contribution, parametrised by the total width~\cite{Isgur:1988vm,Kuhn:1990ad}.
This total width also includes decays to $K^*K$, giving the fit to $\tau\to\pi2\pi^0$ data extra degrees of freedom.
These different assumptions might explain the large discrepancy in reported $a_1$ mass and width to other measurements and PDG averages. 
The dispersive approach in particular was shown to lead to systematically larger widths~\cite{Kuhn:1990ad}.
For modelling the $a_1$ width we use Eq.(\ref{eq:basic_bw}), including $\rho$ and $\rho^\prime$ contributions in the running width with relative weight of -0.145, consistent with a frequently employed~\cite{Jadach:1990mz, Hagiwara:2012vz} prescription by Kuhn and Santamaria~\cite{Kuhn:1990ad}.

\begin{table}[ht]
    \centering
    \resizebox{.49\textwidth}{!}{
    \begin{tabular}{l l|cc|cc}
    \hline
    \multicolumn{3}{c}{ARGUS~\cite{ARGUS:1992olh}}&\multicolumn{3}{c}{$a_1^\pm\to\pi^\mp\pi^\pm\pi^\pm$}\\
    \hline
       & & $m^0_{a_1}$\,[MeV]& $\Gamma^0_{a_1}$\,[MeV] & & \\
       & & 1211$^{+50}_{-7}$ & 446$^{+142}_{-21}$ & & \\
    \hline
        $R$ & $L$ & $m^0_R $\,[MeV] & $\Gamma^0_R$\,[MeV] & \multicolumn{2}{c}{$A_R$} \\
    \hline
        $\rho$ & S & \multirow{2}{*}{773} & \multirow{2}{*}{143} & \multicolumn{2}{c}{1}    \\
        $\rho$ & D &                      &                     & \multicolumn{2}{c}{-0.11$\pm$0.02}  \\
    \hline\\[.1em]
    \hline
    \multicolumn{3}{c}{CLEO~\cite{CLEO:1999rzk}}&\multicolumn{3}{c}{$a_1^\pm\to\pi^\pm\pi^0\pi^0$}\\
    \hline
       & & $m^0_{a_1}$\,[MeV]& $\Gamma^0_{a_1}$\,[GeV] & & \\
       & &  1331$\pm$10 & 814$\pm$10 & & \\
    \hline
        $R$ & $L$ & $m^0_R $\,[MeV] & $\Gamma^0_R$\,[MeV] & $|A_R|$ & $\phi_{A_R}/\pi$\\
    \hline
        \multirow{2}{*}{$\rho$} & S & \multirow{2}{*}{774} & \multirow{2}{*}{149} & 1   & 0 \\
         & D &                        &                        & 0.37$\pm$0.09 & -0.15$\pm$0.10 \\
        \multirow{2}{*}{$\rho^\prime$} & S & \multirow{2}{*}{1370} & \multirow{2}{*}{386} & 0.12$\pm$0.09   & 0.99$\pm$0.25 \\
        & D &                        &                        & 0.87$\pm$0.30 & 0.53$\pm$0.17 \\
        $f_0(500)$  & P & 860  & 880 & 2.10$\pm$0.28 & 0.23$\pm$0.04 \\
        $f_2(1270)$ & P & 1275 & 185& 0.71$\pm$0.17 & 0.56$\pm$0.10 \\
        $f_0(1370)$ & P & 1186 & 350 & 0.77$\pm$0.15 & -0.54$\pm$0.06 \\
    \hline
    \end{tabular}
    }
    \caption{Parameters for modelling the $a_1$ decays as given by $\tau$ decay measurements~\cite{ARGUS:1992olh,CLEO:1999rzk}.
    The respective fits by the experiments resulted in the BW mass $m^0_{a_1}$, width $\Gamma^0_{a_1}$, and complex parameters $\beta$ of intermediate resonances $R$ taken to have BW mass $m^0_R$ and width $\Gamma^0_R$.}
    \label{tab:a1_parameters}
\end{table}

Notably, the CLEO measurement also gives significant coupling to scalar mediators through P wave contributions. 
More recent preliminary results by Belle seem to suggest that the scalar resonances could indeed play a major role~\cite{Rabusov:2023tna}.
This is relevant to the HNL case, not only due to the differential shape in all $a_1$ mediated decays, but also because the Neutral Current mediated decay $N\to\nu3\pi^0$ might be mediated in the chain $N\to\nu(a_1\to((f_0\to2\pi^0)\pi^0))$, which is forbidden in the $\rho$ mediated model due to G-parity.

Given that the resulting CLEO fit parameters of the intermediate resonances are in conflict with PDG values~\cite{ParticleDataGroup:2022pth}, we decide to only use the simpler $\rho$ meson parametrisation described by ARGUS also in the neutral case.
This parametrisation yields estimates for the $\tau$ width comparable to dedicated $\tau$ MC generators and PDG averages (see Table~\ref{tab:tau_widths}).
Differences to \textsc{TauDecay} and an old version of \textsc{Tauola} relying on similar expansions are mainly due to the inclusion of $\rho$ $D$-waves.

\begin{table}[ht]
    \centering
    \begin{tabular}{c|ccc|c}
        \hline
        $H$ & \textsc{Tauola} & \textsc{TauDecay} & this work & PDG\\
        \hline
         $3\pi$     & 2.21 & 2.22 & 2.07 & 2.04\\ 
         $\pi2\pi^0$& 2.25 & 2.27 & 2.12 & 2.10 \\
        \hline
    \end{tabular}
    \caption{Partial width of the $\tau$ lepton to 3 pion states in $10^{-13}$\,GeV. 
    Comparison to existing $\tau$ MC generators~\cite{Jadach:1990mz,Hagiwara:2012vz} and PDG averages~\cite{PDG2026}.
    }\label{tab:tau_widths}
\end{table}
We find that this parametrisation also describes the CLEO data well enough at a differential level for the purposes of this work as shown in Fig.~\ref{fig:cleo_3pi_distribution}, which compares the resulting invariant mass distribution measured by the CLEO~\cite{CLEO:1999rzk} and ALEPH~\cite{ALEPH:2005qgp} collaborations with that expected from the only $\rho$ parametrisation employed in this work. 
Notably, our result agrees with ALEPH data for $m_{\pi2\pi^0}>1.6$GeV, while CLEO suggests a larger differential width.
For $m_{\pi2\pi^0}<0.8$GeV, however, the opposite is true, showcasing the difficulty in normalising the measurement at the kinematic tail ends.
Pending conclusive results from the Belle partial wave analysis, we will therefore resort to the simple $\rho$ mediated model.

\begin{figure}
    \centering
    \includegraphics[width=\linewidth]{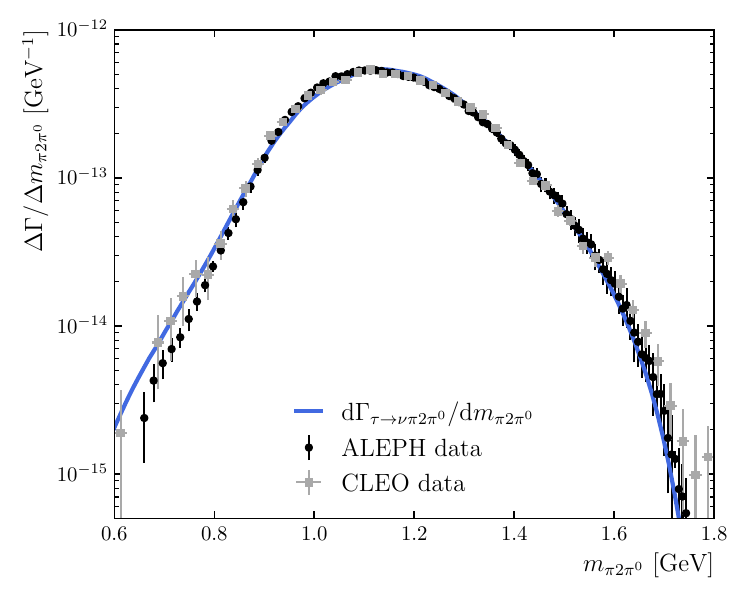}
    \caption{Normalised number of $\tau\to \nu \pi2\pi^0$ events in bins of $\pi2\pi^0$ invariant mass observed by CLEO~\cite{CLEO:1999rzk} and expected from the parametrisation applied in this work.} 
    \label{fig:cleo_3pi_distribution}
\end{figure}

\subsection{Massive quark channels}\label{sec:massive_quark_channels}

\begin{table}[ht]
    \centering
    \begin{tabular}{c|cc}
    \hline
        $q\bar q$   & $h_P(m_h\,[\mathrm{MeV}])$ & $h_V(m_h\,[\mathrm{MeV}])$ \\
        \hline
        $s\bar s$   & $\eta^\prime(958)$& $\phi(1020)$	    \\
        $c\bar c$   & $\eta_c(2984)$    & $J/\Psi(3097)$   	\\ 
        $b\bar b$   & $\eta_b(9460)$    & $\Upsilon(9460)$	\\
        $c\bar d$   & $\eta_b(9460)$    & $D^*(2010)$	    \\
        $c\bar s$   & $D(1968)$         & $D_s^*(2112)$	    \\
        $u\bar b$   & $B^+(5279)$       & $B^*(5325)$	    \\
        $c\bar b$   & $B_c(6274)$       & $B_c^*(6339)$	    \\
    \hline
    \end{tabular}
    \caption{Pseudo-scalar ($h_P$) and vector resonances ($h_P$) considered in this work to account for the resonant part $0<q<q^\mathrm{Trans}_Q$ of a given $q\bar q$ channel. 
    Masses are taken from Ref.s~\cite{PDG2026,CMS:2025byz,ATLAS:2026ubk}.}
    \label{tab:heavy_two_bodies}
\end{table}

For calculating the total widths of heavy neutral leptons with masses above 3 GeV, the quark masses of charm quarks $m_c$ (and also beauty $m_b$) become relevant. 
This is mostly the case through the $c\bar s$ and $c\bar c$ channels.
In $\overline{\mathrm{MS}}$ scheme, lattice QCD lists these the quark masses evaluated at scale $\mu$ as $m_s(2\, \mathrm{GeV}) =$0.924, $m_c(m_c)=$1.280 $m_b(m_b)=$4.171~\cite{FLAG:2024oxs}.
However, as shown in Table~\ref{tab:meson_quark_mass_relation}, the thus derived quark masses show no clear relation to the masses of the resonances they form.
Here we account for QCD running of the $\overline{\mathrm{MS}}$ masses in the QCD renormalisation group (RNG) evaluating the quark pair mass at the resonance mass ($\mu=m_R$).

Setting instead $m_s=$0.15, $m_c=$1.2, and $m_b=$4.35 as approximate physical pole masses and disregarding their RNG running, we find a relationship $m_R-\sum{m_q}\simeq$0.7 to 0.8\,GeV.
This is consistent with the location of the $\rho$ and $K^*$ poles in the light quark cases. 
The only exception to this is the recently measured $B^*$ resonance~\cite{CMS:2025byz}, where the difference is $0.975$\,GeV.
However, this channel only contributes negligibly to the total HNL width, due to a $\lambda^3$ Cabbibo suppression (where $\lambda$ is the expansion parameter of the Wolfenstein CKM parametrisation).
For the purposes of this work the quark masses $m_d$ and $m_u$ might be set to zero, but will be kept at 7.6 and 3.5\,MeV respectively, consistent with the ratio to $m_s$.

The vector states considered in the two body decays of massive quark channels here are fairly narrow, the broadest of which is the $\phi$ meson with $\Gamma_\phi =4.2$\,MeV.
By accounting for the partial HNL widths from massive quark channels only for $q_H>q^\mathrm{trans}_{H,\mathrm{heavy}}-\sum_Q\simeq1$\,GeV, we forego double counting these widths.
In practise, varying this parameter in the range from $0.7$ to $1.25$ only impacts the total width evaluation at the \% level.

\begin{table}[ht]
    \centering
    \begin{tabular}{cc|ccc}
    \hline
        quarks      & resonance & $m_R-\sum{m_q}$ ($\overline{\mathrm{MS}}$) & $m_R-\sum{m_q}$(pole) \\
        \hline
        $s\bar s$   & $\phi(1020)$	    & 0.777 & 0.720 \\
        $c\bar c$   & $J/\Psi(3097)$   	& 1.139 & 0.697 \\ 
        $b\bar b$   & $\Upsilon(9460)$	& 2.206 & 0.760 \\
        $c\bar d$   & $D^*(2010)$	    & 0.912 & 0.810 \\
        $c\bar s$   & $D_s^*(2112)$	    & 0.944 & 0.762 \\
        $c\bar b$   & $B_c^*(6339)$	    & 1.618 & 0.788 \\
        $u\bar b$   & $B^*(5325)$	    & 1.339 & 0.975 \\
    \hline
    \end{tabular}
    \caption{Difference between and sum of quark masses and the lowest lying vector resonance constituting these quark pairs.}
    \label{tab:meson_quark_mass_relation}
\end{table}

\section{Tau decay structure}

In order to match hadronisation rates, in particular those decay channels with large differential widths at the transition point would be relevant for comparison. 
Table~\ref{tab:tau_decays} lists relevant $\tau$ semileptonic decay channels, and shows which of these would be relevant under the assumption of a flat $q_H$ dependence of the differential width. 
These turn out to be completely equivalent, as $\tau$ decay final states including kaons are severely suppressed with respect to those containing only pions to begin with.
The measure further underestimates that multi-hadron states are typically mediated by heavier resonances and, thus, are concentrated at large $q_H$.
Comparing this Table to Fig.~\ref{fig:tau_hadronisation_ratio}, it is clear that inferring the importance of HNL decay channels from their corresponding $\tau$ decay channels branching ratio needs to be handled with care.

\begin{table}[ht]
    \centering
    \begin{tabular}{c|c|c}
        $H$ & $\mathrm{Br}(\tau\to\nu H)$ [\%] & $\frac{\mathrm{Br}(\tau\to\nu H)m_\tau}{m_\tau-q^{\min}_H}\times100$\\[5pt]
        \hline
         $\pi$ & 10.82 & 11.74\\    
         $K$ & 0.70 & 0.97\\
        \hline
         $\pi\pi^0$ & 25.49 & 30.15 \\
         $\pi K^0$ & 0.89 & 1.39 \\
         $K\pi^0$ & 0.43 & 0.67 \\
         $\pi\omega$ & 0.20 &  0.42 \\
         $KK^0$ & 0.15 & 0.34\\
        \hline
         $\pi2\pi^0$ & 9.26 &  12.03\\
         $3\pi$ & 8.99 & 11.76\\
         $\pi K^0\pi^0$ & 0.38 & 0.67 \\
         $K2\pi$ & 0.29 & 0.51 \\
         $KK^0\pi^0$ & 0.15 & 0.41 \\
         $\pi2K^0$ & 0.11 & 0.3 \\
         $\pi\pi^0\eta$ & 0.14 & 0.26\\
        \hline
         $3\pi\pi^0$ & 2.74 & 3.98\\
         $\pi3\pi^0$ & 1.04 & 1.50\\
    \end{tabular}
    \caption{Hadronic $\tau$ decay modes ($\tau\to \nu H$) with branching ratios greater than 0.1\,\%.\cite{PDG2026}, and branching ratios scaled with the inverse available phase space }
    \label{tab:tau_decays}
\end{table}

\section{SHiP geometry and cuts}\label{sec:SHiP_cuts}
    Since its first inception, the concept design of the SHiP experiment has undergone major changes.
    Here, we aim to implement the experiment as given in the BDF/SHiP technical proposal.~\cite{SHiP2023}.
    A comparison between \textsc{Alpinist} and sensitivities determined with the FairShip framework is available in Ref.~\cite{Schubert:2024hpm}.
    
    At the SHiP experiment, it is foreseen to have 400\,GeV SPS protons imping on a Molybdenium target.
    A 50\,m long decay volume begins 33.5\,m downstream of the target front face (vacuum in proposal, now likely filled with He).
    Important components of the SHiP experiment are a $4\times 6 \,\mathrm{m^2}$ spectrometer system just behind the decay volume (DV), featuring a 162.5\,mT$\times 4$\,m magnet.
    Further downstream, the ${5.3\times 6.7}\,\mathrm{m^2}$ electromagnetic calorimeter, and a subsequent ${5.2 \times 6.6}\,\mathrm{m^2}$ muon/hadronic calorimeter are foreseen. 
    This results in a 3.9\,msr  solid angle coverage as seen from the front face of the target.
    
    Beyond the resulting geometrical cuts, we use the prescribed signal cuts on final state kinematics~\cite{SHiP2023}
    \begin{enumerate}
        \setlength\itemsep{0em}
        \item 1\,GeV$c^{-1}< p_\mathrm{track}$,
        \item $\min |\Vec{r}_\mathrm{vtx2} - \Vec{r}_\mathrm{DV wall}| > 5$\,cm,
    \end{enumerate}
    which demands that all observable final state particles have a momentum greater than 1\,GeV$c^{-1}$ and the reconstructed decay vertex be 5\,cm away from the inner wall of the decay volume.
    These are complemented for decay channels including a neutrino by the additional requirements
    \begin{enumerate}
        \setlength\itemsep{0em}
        \setcounter{enumi}{2}
        \item $ z_\mathrm{vtx2} - z_\mathrm{DV} > 1$\,m,
        \item $b_\mathrm{target} < 2.5$\,m.
    \end{enumerate}
    Here, $b_\mathrm{target}$ is the reconstructed LLP track's impact parameter with respect to the beam axis at the interaction point.
    The track reconstruction efficiency is approximated as 100\,\%.

\bibliographystyle{spphys_improved}       % APS-like style for physics
\bibliography{bibliography}

@techreport{SHiP2023,
      author        = "Albanese, R and others",
      collaboration = "SHiP",
      title         = "{BDF/SHiP at the ECN3 high-intensity beam facility}",
      institution   = "CERN",
      reportNumber  = "CERN-SPSC-2023-033, SPSC-P-369",
      address       = "Geneva",
      year          = "2023"
}

@article{Balkin:2025enj,
    author = "Balkin, Reuven and Coren, Ta'el and Soreq, Yotam and Williams, Mike",
    title = "{A covariant description of the interactions of axion-like particles and hadrons}",
    eprint = "2506.15637",
    archivePrefix = "arXiv",
    primaryClass = "hep-ph",
    doi = "10.1007/JHEP03(2026)086",
    journal = "JHEP",
    volume = "03",
    pages = "086",
    year = "2026"
}

@article{Jerhot:2022chi,
    author = {Jerhot, Jan  and others},
    title = "{ALPINIST: Axion-Like Particles In Numerous Interactions Simulated and Tabulated}",
    eprint = "2201.05170",
    archivePrefix = "arXiv",
    primaryClass = "hep-ph",
    reportNumber = "TTK-22-04, CP3-22-02",
    doi = "10.1007/JHEP07(2022)094",
    journal = "JHEP",
    volume = "07",
    pages = "094",
    year = "2022"
}

@techreport{deBlas:2025gyz,
    author = "de Blas, Jorge and others",
    title = "{Physics Briefing Book: Input for the 2026 update of the European Strategy for Particle Physics}",
    institution="CERN",
    eprint = "2511.03883",
    archivePrefix = "arXiv",
    primaryClass = "hep-ex",
    reportNumber = "CERN--2025-008, CERN-ESU-2025-001",
    doi = "10.23731/CYRM-2025-008",
    month = "11",
    year = "2025"
}

@techreport{SHiP:2025ows,
    author = "Albanese, R. and others",
    group = "SHiP, HI-ECN3 Project Team",
    title = "{SHiP experiment at the SPS Beam Dump Facility}",
    institution ="CERN" ,
    eprint = "2504.06692",
    archivePrefix = "arXiv",
    primaryClass = "hep-ex",
    month = "4",
    year = "2025"
}

@misc{Alpinist_Zenodo,
  author       = { D\"{o}brich, B. and Ertas, E. Jerhot, J. and and Kahlhoefer, F. and Schubert, J. and Spadaro, T.},
  title        = {{ALPINIST}: v1.5},
  month        = sep,
  year         = 2026,
  publisher    = {Zenodo},
  version      = {v1.5},
  doi          = {10.5281/zenodo.13167444},
  note         = {10.5281/zenodo.13167444}
}

@article{Schubert:2024hpm,
    author = {Schubert, Jonathan L. and D{\"o}brich, Babette and Jerhot, Jan and Spadaro, Tommaso},
    title = "{On the impact of heavy meson production spectra on searches for heavy neutral leptons}",
    eprint = "2407.08673",
    archivePrefix = "arXiv",
    primaryClass = "hep-ph",
    reportNumber = "MPP-2024-129",
    doi = "10.1007/JHEP02(2025)140",
    journal = "JHEP",
    volume = "02",
    pages = "140",
    year = "2025"
}

@article{Blackstone:2024ouf,
    author = "Blackstone, Patrick J. and Tarr{\'u}s Castell{\`a}, Jaume and Passemar, Emilie and Zupan, Jure",
    title = "{Hadronic Decays of a Higgs-mixed Scalar}",
    eprint = "2407.13587",
    archivePrefix = "arXiv",
    primaryClass = "hep-ph",
    month = "7",
    year = "2024"
}

@article{Gieseke:2025gfq,
    author = "Gieseke, Stefan and Kahlhoefer, Felix and Seebach, Henry",
    title = "{Branching ratios for Higgs-mixed scalars at the GeV scale from hadronisation models with conservation laws}",
    eprint = "2510.17961",
    archivePrefix = "arXiv",
    primaryClass = "hep-ph",
    reportNumber = "KA-TP-33-2025, TTP25-037, P3H-25-078",
    doi = "10.1016/j.physletb.2026.140321",
    journal = "Phys. Lett. B",
    volume = "875",
    pages = "140321",
    year = "2026"
}

@techreport{Bellm:2025pcw,
      author        = "Bellm, J. and others",
      title         = "{The Physics of Herwig 7}",
      archivePrefix = "arXiv",
      eprint        = "2512.16645",
      reportNumber  = "CERN-TH-2025-252, IPPP-25-57, HERWIG-2025-01,
                       KA-TP-36-2025, MCNET-25-31",
      year          = "2025",
      institution   = "CERN"
}

@article{Bierlich:2022pfr,
    author = "Bierlich, Christian and others",
    title = "{A comprehensive guide to the physics and usage of PYTHIA 8.3}",
    eprint = "2203.11601",
    archivePrefix = "arXiv",
    primaryClass = "hep-ph",
    reportNumber = "LU-TP 22-16, MCNET-22-04, FERMILAB-PUB-22-227-SCD",
    doi = "10.21468/SciPostPhysCodeb.8",
    journal = "SciPost Phys. Codeb.",
    volume = "2022",
    pages = "8",
    year = "2022"
}

@misc{SensCalc,
  author       = {Maksym Ovchynnikov},
  title        = {SensCalc},
  month        = feb,
  year         = 2024,
  publisher    = {Zenodo},
  version      = {1.1},
  doi          = {10.5281/zenodo.10672537}
}

@inproceedings{Batell:2022dpx,
    author = "Batell, Brian and Blinov, Nikita and Hearty, Christopher and McGehee, Robert",
    title = "{Exploring Dark Sector Portals with High Intensity Experiments}",
    booktitle = "{Snowmass 2021}",
    eprint = "2207.06905",
    archivePrefix = "arXiv",
    primaryClass = "hep-ph",
    month = "7",
    year = "2022"
}

@techreport{PBC:2025sny,
    author = "Alemany Fern{\'a}ndez, R. and others",
    collaboration = "PBC",
    institution   = "CERN",
    title = "{Summary Report of the Physics Beyond Colliders Study at CERN}",
    eprint = "2505.00947",
    archivePrefix = "arXiv",
    primaryClass = "hep-ex",
    reportNumber = "CERN-PBC-REPORT-2025-003",
    
    month = "5",
    year = "2025"
}

@article{Kretz:2025pfk,
    author = "Kretz, Tim and Nierste, Ulrich",
    title = "{QCD corrections to charged-current decays with Heavy Sterile Neutrinos in initial or final state and their impact on {\ensuremath{\tau}} decays}",
    eprint = "2512.00476",
    archivePrefix = "arXiv",
    primaryClass = "hep-ph",
    reportNumber = "P3H--25--103, TTP25-049",
    doi = "10.1007/JHEP06(2026)146",
    journal = "JHEP",
    volume = "06",
    pages = "146",
    year = "2026"
}

@article{Groote:2013xt,
    author = "Groote, S. and Korner, J. G. and Tuvike, P.",
    title = "{Fully analytical ${\cal O}(\alpha_s)$ results for on-shell and off-shell polarized W-boson decays into massive quark pairs}",
    eprint = "1301.0881",
    archivePrefix = "arXiv",
    primaryClass = "hep-ph",
    reportNumber = "MZ-TH-12-55",
    doi = "10.1140/epjc/s10052-013-2454-2",
    journal = "Eur. Phys. J. C",
    volume = "73",
    pages = "2454",
    year = "2013"
}

@article{FLAG:2024oxs,
    author = "Aoki, Y. and others",
    collaboration = "Flavour Lattice Averaging Group (FLAG)",
    title = "{FLAG review 2024}",
    eprint = "2411.04268",
    archivePrefix = "arXiv",
    primaryClass = "hep-lat",
    reportNumber = "CERN-TH-2024-192, FERMILAB-PUB-24-0785-T",
    doi = "10.1103/nfzp-p5dn",
    journal = "Phys. Rev. D",
    volume = "113",
    number = "1",
    pages = "014508",
    year = "2026"
}

@article{Chen:2020qma,
    author = "Chen, Ying and Chiu, Wei-Feng and Gong, Ming and Liu, Zhaofeng and Ma, Yunheng",
    collaboration = "{\ensuremath{\chi}}QCD",
    title = "{Charmed and $\phi$ meson decay constants from 2+1-flavor lattice QCD}",
    eprint = "2008.05208",
    archivePrefix = "arXiv",
    primaryClass = "hep-lat",
    doi = "10.1088/1674-1137/abcd8f",
    journal = "Chin. Phys. C",
    volume = "45",
    number = "2",
    pages = "023109",
    year = "2021"
}

@article{CMS:2025byz,
    author = "Hayrapetyan, Aram and others",
    collaboration = "CMS",
    title = "{First Exclusive Reconstruction of the B*+, B*0, and Bs*0 Mesons and Precise Measurement of Their Masses}",
    eprint = "2508.05820",
    archivePrefix = "arXiv",
    primaryClass = "hep-ex",
    reportNumber = "CMS-BPH-24-011, CERN-EP-2025-162",
    doi = "10.1103/njf9-4zfv",
    journal = "Phys. Rev. Lett.",
    volume = "136",
    number = "3",
    pages = "031902",
    year = "2026"
}

@article{ATLAS:2026ubk,
    author = "Aad, Georges and others",
    collaboration = "ATLAS",
    title = "{Observation of a $B_c^{*+}$ meson with the ATLAS detector}",
    eprint = "2605.16228",
    archivePrefix = "arXiv",
    primaryClass = "hep-ex",
    reportNumber = "CERN-EP-2026-142",
    month = "5",
    year = "2026"
}

@article{Tao:2022qxa,
    author = "Tao, Wei and Zhu, Ruilin and Xiao, Zhen-Jun",
    title = "{Next-to-next-to-leading order matching of beauty-charmed meson Bc and Bc* decay constants}",
    eprint = "2209.15521",
    archivePrefix = "arXiv",
    primaryClass = "hep-ph",
    doi = "10.1103/PhysRevD.106.114037",
    journal = "Phys. Rev. D",
    volume = "106",
    number = "11",
    pages = "114037",
    year = "2022"
}

@article{Davies:2010ip,
    author = "Davies, C. T. H. and McNeile, C. and Follana, E. and Lepage, G. P. and Na, H. and Shigemitsu, J.",
    title = "{Update: Precision $D_s$ decay constant from full lattice QCD using very fine lattices}",
    eprint = "1008.4018",
    archivePrefix = "arXiv",
    primaryClass = "hep-lat",
    doi = "10.1103/PhysRevD.82.114504",
    journal = "Phys. Rev. D",
    volume = "82",
    pages = "114504",
    year = "2010"
}

@article{Hatton:2021dvg,
    author = "Hatton, D. and Davies, C. T. H. and Koponen, J. and Lepage, G. P. and Lytle, A. T.",
    title = "{Bottomonium precision tests from full lattice QCD: Hyperfine splitting, \ensuremath{\upsilon} leptonic width, and b quark contribution $\ensuremath{e^+e^- \to}$ hadrons}",
    eprint = "2101.08103",
    archivePrefix = "arXiv",
    primaryClass = "hep-lat",
    doi = "10.1103/PhysRevD.103.054512",
    journal = "Phys. Rev. D",
    volume = "103",
    number = "5",
    pages = "054512",
    year = "2021"
}

@article{Pivovarov:2000cr,
    author = "Pivovarov, A. A.",
    title = "{Running electromagnetic coupling constant: Low-energy normalization and the value at M(Z)}",
    eprint = "hep-ph/0011135",
    archivePrefix = "arXiv",
    reportNumber = "MZ-TH-00-51",
    doi = "10.1134/1.1495645",
    journal = "Phys. Atom. Nucl.",
    volume = "65",
    pages = "1319--1340",
    year = "2002"
}

@article{Feldmann:1999uf,
    author = "Feldmann, Thorsten",
    title = "{Quark structure of pseudoscalar mesons}",
    eprint = "hep-ph/9907491",
    archivePrefix = "arXiv",
    reportNumber = "WUB-99-19",
    doi = "10.1142/S0217751X00000082",
    journal = "Int. J. Mod. Phys. A",
    volume = "15",
    pages = "159--207",
    year = "2000"
}

@article{Feldmann:1998vh,
    author = "Feldmann, T. and Kroll, P. and Stech, B.",
    title = "{Mixing and decay constants of pseudoscalar mesons}",
    eprint = "hep-ph/9802409",
    archivePrefix = "arXiv",
    reportNumber = "WU-B-98-2, HD-THEP-98-5",
    doi = "10.1103/PhysRevD.58.114006",
    journal = "Phys. Rev. D",
    volume = "58",
    pages = "114006",
    year = "1998"
}

@article{Ottnad:2025zxq,
    author = "Ottnad, Konstantin and Bacchio, Simone and Finkenrath, Jacob and Kostrzewa, Bartosz and Petschlies, Marcus and Pittler, Ferenc and Urbach, Carsten and Wenger, Urs",
    title = "{$\eta $, $\eta ^\prime $ mesons from lattice QCD in fully physical conditions}",
    eprint = "2503.09895",
    archivePrefix = "arXiv",
    primaryClass = "hep-lat",
    reportNumber = "MITP-25-021, CERN-TH-2025-052",
    doi = "10.1140/epja/s10050-025-01635-0",
    journal = "Eur. Phys. J. A",
    volume = "61",
    number = "7",
    pages = "169",
    year = "2025",
    note = "[Erratum: Eur.Phys.J.A 61, 227 (2025)]"
}

@article{Finkemeier:1996dh,
    author = "Finkemeier, Markus and Mirkes, Erwin",
    title = "{The Scalar contribution to tau --{\ensuremath{>}} k pi tau-neutrino}",
    eprint = "hep-ph/9601275",
    archivePrefix = "arXiv",
    reportNumber = "TTP-95-44, HUTP-95-A050",
    doi = "10.1007/s002880050284",
    journal = "Z. Phys. C",
    volume = "72",
    pages = "619--626",
    year = "1996"
}

@article{Bernard:2011ae,
    author = "Bernard, V. and Boito, D. R. and Passemar, E.",
    editor = "Lafferty, George and Soldner-Rembold, Stefan",
    title = "{Dispersive representation of the scalar and vector $\rm{K}\pi$ form factors for $\tau \to K \pi \nu_\tau$ and $K_{\ell3}$ decays}",
    eprint = "1103.4855",
    archivePrefix = "arXiv",
    primaryClass = "hep-ph",
    reportNumber = "UAB-FT-691, IFIC-11-12",
    doi = "10.1016/j.nuclphysbps.2011.06.024",
    journal = "Nucl. Phys. B Proc. Suppl.",
    volume = "218",
    pages = "140--145",
    year = "2011"
}

@article{Jamin:2006tk,
    author = "Jamin, Matthias and Pich, Antonio and Portoles, Jorge",
    title = "{Spectral distribution for the decay tau ---{\ensuremath{>}} nu(tau) K pi}",
    eprint = "hep-ph/0605096",
    archivePrefix = "arXiv",
    reportNumber = "UAB-FT-601, IFIC-06-10, FTUV-06-0509",
    doi = "10.1016/j.physletb.2006.06.058",
    journal = "Phys. Lett. B",
    volume = "640",
    pages = "176--181",
    year = "2006"
}

@article{Hoferichter:2014vra,
    author = "Hoferichter, Martin and Kubis, Bastian and Leupold, Stefan and Niecknig, Franz and Schneider, Sebastian P.",
    title = "{Dispersive analysis of the pion transition form factor}",
    eprint = "1410.4691",
    archivePrefix = "arXiv",
    primaryClass = "hep-ph",
    doi = "10.1140/epjc/s10052-014-3180-0",
    journal = "Eur. Phys. J. C",
    volume = "74",
    pages = "3180",
    year = "2014"
}

@article{Dhiman:2017urn,
    author = "Dhiman, Nisha and Dahiya, Harleen",
    title = "{Decay constants of pseudoscalar and vector $B$ and $D$ mesons in the light-cone quark model}",
    eprint = "1708.07274",
    archivePrefix = "arXiv",
    primaryClass = "hep-ph",
    doi = "10.1140/epjp/i2018-11972-5",
    journal = "Eur. Phys. J. Plus",
    volume = "133",
    number = "4",
    pages = "134",
    year = "2018"
}

@article{Belle:2008xpe,
    author = "Fujikawa, M. and others",
    collaboration = "Belle",
    title = "{High-Statistics Study of the tau- ---{\ensuremath{>}} pi- pi0 nu(tau) Decay}",
    eprint = "0805.3773",
    archivePrefix = "arXiv",
    primaryClass = "hep-ex",
    reportNumber = "BELLE-2008-16, KEK-2008-10",
    doi = "10.1103/PhysRevD.78.072006",
    journal = "Phys. Rev. D",
    volume = "78",
    pages = "072006",
    year = "2008"
}

@article{BaBar:2012bdw,
    author = "Lees, J. P. and others",
    collaboration = "BaBar",
    title = "{Precise Measurement of the $e^+ e^- \to \pi^+\pi^- (\gamma)$ Cross Section with the Initial-State Radiation Method at BABAR}",
    eprint = "1205.2228",
    archivePrefix = "arXiv",
    primaryClass = "hep-ex",
    reportNumber = "BABAR-PUB-12-003",
    doi = "10.1103/PhysRevD.86.032013",
    journal = "Phys. Rev. D",
    volume = "86",
    pages = "032013",
    year = "2012"
}

@article{Gonzalez-Solis:2019iod,
    author = "Gonz{\`a}lez-Sol{\'\i}s, Sergi and Roig, Pablo",
    title = "{A dispersive analysis of the pion vector form factor and $\tau ^{-}\rightarrow K^{-}K_{S}\nu _{\tau }$ decay}",
    eprint = "1902.02273",
    archivePrefix = "arXiv",
    primaryClass = "hep-ph",
    doi = "10.1140/epjc/s10052-019-6943-9",
    journal = "Eur. Phys. J. C",
    volume = "79",
    number = "5",
    pages = "436",
    year = "2019"
}

@article{Stamen:2022uqh,
    author = "Stamen, Dominik and Hariharan, Deepti and Hoferichter, Martin and Kubis, Bastian and Stoffer, Peter",
    title = "{Kaon electromagnetic form factors in dispersion theory}",
    eprint = "2202.11106",
    archivePrefix = "arXiv",
    primaryClass = "hep-ph",
    reportNumber = "PSI-PR-22-04, ZU-TH 22/22",
    doi = "10.1140/epjc/s10052-022-10348-3",
    journal = "Eur. Phys. J. C",
    volume = "82",
    number = "5",
    pages = "432",
    year = "2022"
}

@article{BaBar:2018qry,
    author = "Lees, J. P. and others",
    collaboration = "BaBar",
    title = "{Measurement of the spectral function for the $\tau^-\to K^-K_S\nu_{\tau}$ decay}",
    eprint = "1806.10280",
    archivePrefix = "arXiv",
    primaryClass = "hep-ex",
    reportNumber = "BABAR-PUB-18/005, SLAC-PUB-17286, BABAR-PUB-18-005",
    doi = "10.1103/PhysRevD.98.032010",
    journal = "Phys. Rev. D",
    volume = "98",
    number = "3",
    pages = "032010",
    year = "2018"
}

@article{Belle:2007goc,
    author = "Epifanov, D. and others",
    collaboration = "Belle",
    title = "{Study of tau- ---{\ensuremath{>}} K(S) pi- nu(tau) decay at Belle}",
    eprint = "0706.2231",
    archivePrefix = "arXiv",
    primaryClass = "hep-ex",
    reportNumber = "BELLE-PREPRINT-2007-28, KEK-PREPRINT-2007-17",
    doi = "10.1016/j.physletb.2007.08.045",
    journal = "Phys. Lett. B",
    volume = "654",
    pages = "65--73",
    year = "2007"
}

@article{Deur:2023dzc,
    author = "Deur, Alexandre and Brodsky, Stanley J. and Roberts, Craig D.",
    title = "{QCD running couplings and effective charges}",
    eprint = "2303.00723",
    archivePrefix = "arXiv",
    primaryClass = "hep-ph",
    reportNumber = "JLAB-PHY-23-3758, SLAC--PUB--17724, NJU-INP 071/23",
    doi = "10.1016/j.ppnp.2023.104081",
    journal = "Prog. Part. Nucl. Phys.",
    volume = "134",
    pages = "104081",
    year = "2024"
}

@article{Baikov:2008jh,
    author = "Baikov, P. A. and Chetyrkin, K. G. and Kuhn, Johann H.",
    title = "{Order alpha**4(s) QCD Corrections to Z and tau Decays}",
    eprint = "0801.1821",
    archivePrefix = "arXiv",
    primaryClass = "hep-ph",
    reportNumber = "SFB-CPP-08-04, TTP08-01",
    doi = "10.1103/PhysRevLett.101.012002",
    journal = "Phys. Rev. Lett.",
    volume = "101",
    pages = "012002",
    year = "2008"
}

@article{Gorishnii:1990vf,
    author = "Gorishnii, S. G. and Kataev, A. L. and Larin, S. A.",
    title = "{The $O(\alpha^{3}_{s})$-corrections to $\sigma_{tot}(e^{+}e^{-}\rightarrow hadrons)$ and $\Gamma(\tau^{-} \rightarrow \nu_{\tau} + hadrons)$ in QCD}",
    reportNumber = "UM-TH-91-01",
    doi = "10.1016/0370-2693(91)90149-K",
    journal = "Phys. Lett. B",
    volume = "259",
    pages = "144--150",
    year = "1991"
}

@article{Isgur:1988vm,
    author = "Isgur, Nathan and Morningstar, Colin and Reader, Cathy",
    title = "{The a1 in tau Decay}",
    reportNumber = "Print-88-0442 (TORONTO)",
    doi = "10.1103/PhysRevD.39.1357",
    journal = "Phys. Rev. D",
    volume = "39",
    pages = "1357",
    year = "1989"
}

@article{Kuhn:1990ad,
    author = "Kuhn, Johann H. and Santamaria, A.",
    title = "{Tau decays to pions}",
    reportNumber = "MPI-PAE/PTh-17/90",
    doi = "10.1007/BF01572024",
    journal = "Z. Phys. C",
    volume = "48",
    pages = "445--452",
    year = "1990"
}

@article{Kuhn:1992nz,
    author = "Kuhn, Johann H. and Mirkes, E.",
    title = "{Structure functions in tau decays}",
    reportNumber = "TTP-92-20",
    doi = "10.1007/BF01474741",
    journal = "Z. Phys. C",
    volume = "56",
    pages = "661--672",
    year = "1992",
    note = "[Erratum: Z.Phys.C 67, 364 (1995)]"
}

@article{Krinner:2021let,
    author = "Krinner, Fabian and Paul, Stephan",
    title = "{Hadronic currents and form factors in three-body semileptonic $\tau $ decays}",
    eprint = "2107.04295",
    archivePrefix = "arXiv",
    primaryClass = "hep-ph",
    doi = "10.1140/epjc/s10052-021-09876-1",
    journal = "Eur. Phys. J. C",
    volume = "81",
    number = "12",
    pages = "1073",
    year = "2021"
}

@article{Jadach:1990mz,
    author = "Jadach, Stanislaw and Kuhn, Johann H. and Was, Zbigniew",
    title = "{TAUOLA: A Library of Monte Carlo programs to simulate decays of polarized tau leptons}",
    reportNumber = "CERN-TH-5856-90",
    doi = "10.1016/0010-4655(91)90038-M",
    journal = "Comput. Phys. Commun.",
    volume = "64",
    pages = "275--299",
    year = "1990"
}

@article{Hagiwara:2012vz,
    author = "Hagiwara, Kaoru and Li, Tong and Mawatari, Kentarou and Nakamura, Junya",
    title = "{TauDecay: a library to simulate polarized tau decays via FeynRules and MadGraph5}",
    eprint = "1212.6247",
    archivePrefix = "arXiv",
    primaryClass = "hep-ph",
    doi = "10.1140/epjc/s10052-013-2489-4",
    journal = "Eur. Phys. J. C",
    volume = "73",
    pages = "2489",
    year = "2013"
}

@article{CLEO:1999rzk,
    author = "Asner, D. M. and others",
    collaboration = "CLEO",
    title = "{Hadronic structure in the decay {\ensuremath{\tau^-\to\nu_\tau\pi^-\pi^0\pi^0}} and the sign of the tau-neutrino helicity}",
    eprint = "hep-ex/9902022",
    archivePrefix = "arXiv",
    reportNumber = "SLAC-REPRINT-1999-097, CLNS-99-1601, CLEO-99-01",
    doi = "10.1103/PhysRevD.61.012002",
    journal = "Phys. Rev. D",
    volume = "61",
    pages = "012002",
    year = "2000"
}

@article{ARGUS:1992olh,
    author = "Albrecht, H. and others",
    collaboration = "ARGUS",
    title = "{Analysis of the decay {\ensuremath{\tau^-\to\pi^-\pi^-\pi^+\nu_\tau}} and determination of the $a_1(1260)$ resonance parameters}",
    reportNumber = "DESY-92-125",
    doi = "10.1007/BF01554080",
    journal = "Z. Phys. C",
    volume = "58",
    pages = "61--70",
    year = "1993"
}

@article{ALEPH:2005qgp,
    author = "Schael, S. and others",
    collaboration = "ALEPH",
    title = "{Branching ratios and spectral functions of tau decays: Final ALEPH measurements and physics implications}",
    eprint = "hep-ex/0506072",
    archivePrefix = "arXiv",
    doi = "10.1016/j.physrep.2005.06.007",
    journal = "Phys. Rept.",
    volume = "421",
    pages = "191--284",
    year = "2005"
}

@article{Rabusov:2023tna,
    author = "Rabusov, Andrei and Greenwald, Daniel and Paul, Stephan",
    collaboration = "Belle",
    title = "{Partial-wave analysis of~$\tau^-\to\pi^-\pi^-\pi^+\nu_\tau$ at BELLE}",
    eprint = "2310.09155",
    archivePrefix = "arXiv",
    primaryClass = "hep-ex",
    doi = "10.1393/ncc/i2024-24154-4",
    journal = "Nuovo Cim. C",
    volume = "47",
    number = "4",
    pages = "154",
    year = "2024"
}

@article{Rabusov:2022woa,
    author = "Rabusov, Andrei and Greenwald, Daniel and Paul, Stephan",
    title = "{Partial wave analysis of $\tau^-\to\pi^-\pi^+\pi^-\nu_\tau$ at Belle}",
    eprint = "2211.11696",
    archivePrefix = "arXiv",
    primaryClass = "hep-ex",
    doi = "10.22323/1.414.1034",
    journal = "PoS",
    volume = "ICHEP2022",
    pages = "1034",
    year = "2022"
}

@article{Finkemeier:1995sr,
    author = "Finkemeier, Markus and Mirkes, Erwin",
    title = "{Tau decays into kaons}",
    eprint = "hep-ph/9503474",
    archivePrefix = "arXiv",
    reportNumber = "MAD-PH-882, LNF-95-015-P",
    doi = "10.1007/s002880050024",
    journal = "Z. Phys. C",
    volume = "69",
    pages = "243--252",
    year = "1996"
}

@article{Czyz:2008kw,
    author = "Czyz, Henryk and Kuhn, Johann H. and Wapienik, Agnieszka",
    title = "{Four-pion production in tau decays and e+e- annihilation: An Update}",
    eprint = "0804.0359",
    archivePrefix = "arXiv",
    primaryClass = "hep-ph",
    doi = "10.1103/PhysRevD.77.114005",
    journal = "Phys. Rev. D",
    volume = "77",
    pages = "114005",
    year = "2008"
}

@article{CLEO:1999heg,
    author = "Edwards, K. W. and others",
    collaboration = "CLEO",
    title = "{Resonant structure of tau ---{\ensuremath{>}} three pi pi0 neutrino(tau) and tau ---{\ensuremath{>}} omega pi neutrino(tau) decays}",
    eprint = "hep-ex/9908024",
    archivePrefix = "arXiv",
    reportNumber = "SLAC-REPRINT-1999-109, CLNS-99-1631, CLEO-99-11",
    doi = "10.1103/PhysRevD.61.072003",
    journal = "Phys. Rev. D",
    volume = "61",
    pages = "072003",
    year = "2000"
}

@article{COMPASS:2020yhb,
    author = "Alexeev, G. D. and others",
    collaboration = "COMPASS",
    title = "{Triangle Singularity as the Origin of the $a_1(1420)$}",
    eprint = "2006.05342",
    archivePrefix = "arXiv",
    primaryClass = "hep-ph",
    reportNumber = "CERN-EP-2020-104",
    doi = "10.1103/PhysRevLett.127.082501",
    journal = "Phys. Rev. Lett.",
    volume = "127",
    number = "8",
    pages = "082501",
    year = "2021"
}

@article{PDG2026,
author = {Takahashi, F. and others},
title = {Review of Particle Physics*},
journal = {International Journal of Modern Physics A},
volume = {41},
number = {22},
pages = {2630011},
year = {2026},
doi = {10.1142/S0217751X26300115},
eprint = { https://doi.org/10.1142/S0217751X26300115}
}

@article{ParticleDataGroup:2022pth,
    author = "Workman, R. L. and others",
    collaboration = "Particle Data Group",
    title = "{Review of Particle Physics}",
    doi = "10.1093/ptep/ptac097",
    journal = "PTEP",
    volume = "2022",
    pages = "083C01",
    year = "2022"
}

@inproceedings{Schubert:2026puf,
    author = "Schubert, Jonathan Leon",
    collaboration = "NA62",
    title = "{Recent Results from NA62 in Kaon and Dump Mode}",
    booktitle = "{60th Rencontres de Moriond on QCD and High Energy Interactions}: {Moriond QCD}",
    eprint = "2605.02415",
    archivePrefix = "arXiv",
    primaryClass = "hep-ex",
    reportNumber = "MPP-2026-83",
    month = "5",
    year = "2026"
}

@article{Feng:2024zfe,
    author = "Feng, Jonathan L. and Hewitt, Alec and Kling, Felix and La Rocco, Daniel",
    title = "{Simulating heavy neutral leptons with general couplings at collider and fixed target experiments}",
    eprint = "2405.07330",
    archivePrefix = "arXiv",
    primaryClass = "hep-ph",
    doi = "10.1103/PhysRevD.110.035029",
    journal = "Phys. Rev. D",
    volume = "110",
    number = "3",
    pages = "035029",
    year = "2024"
}

@article{Dobrich:2026mvi,
    author = {D{\"o}brich, Babette and Jerhot, Jan and Massri, Karim and Schubert, Jonathan L. and Spadaro, Tommaso},
    title = "{NO LESS: Novel Opportunities for Light Exotic Searches at the SPS}",
    eprint = "2601.17119",
    archivePrefix = "arXiv",
    primaryClass = "hep-ex",
    reportNumber = "MPP-2026-4",
    doi = "10.1007/JHEP06(2026)047",
    journal = "JHEP",
    volume = "06",
    pages = "047",
    year = "2026"
}

@inproceedings{Alimena:2025kjv,
    author = "Alimena, J. and others",
    title = "{Feebly-Interacting Particles: FIPs at LHCb {\textemdash} Workshop Report 2025 Edition}",
    booktitle = "{LHCb FIP Physics Workshop 2025}",
    eprint = "2510.05257",
    archivePrefix = "arXiv",
    primaryClass = "hep-ph",
    month = "10",
    year = "2025"
}

@article{CMS:2024ake,
    author = "Hayrapetyan, Aram and others",
    collaboration = "CMS",
    title = "{Search for long-lived heavy neutral leptons decaying in the CMS muon detectors in proton-proton collisions at s=13\,\,TeV}",
    eprint = "2402.18658",
    archivePrefix = "arXiv",
    primaryClass = "hep-ex",
    reportNumber = "CMS-EXO-22-017, CERN-EP-2024-022",
    doi = "10.1103/PhysRevD.110.012004",
    journal = "Phys. Rev. D",
    volume = "110",
    number = "1",
    pages = "012004",
    year = "2024"
}

@article{CMS:2024hik,
    author = "Hayrapetyan, Aram and others",
    collaboration = "CMS",
    title = "{Search for long-lived heavy neutral leptons in proton-proton collision events with a lepton-jet pair associated with a secondary vertex at $ \sqrt{s} $ = 13 TeV}",
    eprint = "2407.10717",
    archivePrefix = "arXiv",
    primaryClass = "hep-ex",
    reportNumber = "CMS-EXO-21-011, CERN-EP-2024-161",
    doi = "10.1007/JHEP02(2025)036",
    journal = "JHEP",
    volume = "02",
    pages = "036",
    year = "2025"
}

@article{ATLAS:2025uah,
    author = "Aad, Georges and others",
    collaboration = "ATLAS",
    title = "{Search for heavy neutral leptons in decays of W bosons using leptonic and semi-leptonic displaced vertices in $ \sqrt{s} $ = 13 TeV pp collisions with the ATLAS detector}",
    eprint = "2503.16213",
    archivePrefix = "arXiv",
    primaryClass = "hep-ex",
    reportNumber = "CERN-EP-2025-052",
    doi = "10.1007/JHEP07(2025)196",
    journal = "JHEP",
    volume = "07",
    pages = "196",
    year = "2025"
}

@article{LHCb:2025ymr,
    author = "Aaij, Roel and others",
    collaboration = "LHCb",
    title = "{Search for heavy neutral leptons in B-meson decays}",
    eprint = "2512.14551",
    archivePrefix = "arXiv",
    primaryClass = "hep-ex",
    reportNumber = "LHCb-PAPER-2025-042, CERN-EP-2025-264",
    doi = "10.1007/JHEP03(2026)178",
    journal = "JHEP",
    volume = "03",
    pages = "178",
    year = "2026"
}

@article{Gorbunov:2007ak,
    author = "Gorbunov, Dmitry and Shaposhnikov, Mikhail",
    title = "{How to find neutral leptons of the $\nu$MSM?}",
    eprint = "0705.1729",
    archivePrefix = "arXiv",
    primaryClass = "hep-ph",
    doi = "10.1088/1126-6708/2007/10/015",
    journal = "JHEP",
    volume = "10",
    pages = "015",
    year = "2007",
    note = "[Erratum: JHEP 11, 101 (2013)]"
}

@article{Atre:2009rg,
    author = "Atre, Anupama and Han, Tao and Pascoli, Silvia and Zhang, Bin",
    title = "{The Search for Heavy Majorana Neutrinos}",
    eprint = "0901.3589",
    archivePrefix = "arXiv",
    primaryClass = "hep-ph",
    reportNumber = "FERMILAB-PUB-08-086-T, NSF-KITP-08-54, MADPH-06-1466, DCPT-07-198, IPPP-07-99",
    doi = "10.1088/1126-6708/2009/05/030",
    journal = "JHEP",
    volume = "05",
    pages = "030",
    year = "2009"
}

@article{Helo:2010cw,
    author = "Helo, Juan Carlos and Kovalenko, Sergey and Schmidt, Ivan",
    title = "{Sterile neutrinos in lepton number and lepton flavor violating decays}",
    eprint = "1005.1607",
    archivePrefix = "arXiv",
    primaryClass = "hep-ph",
    doi = "10.1016/j.nuclphysb.2011.07.020",
    journal = "Nucl. Phys. B",
    volume = "853",
    pages = "80--104",
    year = "2011"
}

@article{Bondarenko:2018ptm,
    author = "Bondarenko, Kyrylo and Boyarsky, Alexey and Gorbunov, Dmitry and Ruchayskiy, Oleg",
    title = "{Phenomenology of GeV-scale Heavy Neutral Leptons}",
    eprint = "1805.08567",
    archivePrefix = "arXiv",
    primaryClass = "hep-ph",
    doi = "10.1007/JHEP11(2018)032",
    journal = "JHEP",
    volume = "11",
    pages = "032",
    year = "2018"
}

@article{Coloma:2020lgy,
    author = "Coloma, Pilar and Fern\'andez-Mart\'\i{}nez, Enrique and Gonz\'alez-L\'opez, Manuel and Hern\'andez-Garc\'\i{}a, Josu and Pavlovic, Zarko",
    title = "{GeV-scale neutrinos: interactions with mesons and DUNE sensitivity}",
    eprint = "2007.03701",
    archivePrefix = "arXiv",
    primaryClass = "hep-ph",
    reportNumber = "FERMILAB-PUB-20-269-ND",
    doi = "10.1140/epjc/s10052-021-08861-y",
    journal = "Eur. Phys. J. C",
    volume = "81",
    number = "1",
    pages = "78",
    year = "2021"
}

\end{document}